\documentclass{aa}  

\usepackage{longtable}
\usepackage{graphicx}
\usepackage{xltabular}
\usepackage{ltablex}
\usepackage{longtable}
\usepackage{multicol}
\usepackage{supertabular}
\usepackage{txfonts}
\usepackage{natbib}
\usepackage{xcolor}
\usepackage{yfonts}

\begin{document}

\title{A confined dynamo: magnetic activity of the K-dwarf component in the pre-cataclysmic binary system V471\,Tauri}

%   \subtitle{}
\titlerunning{Magnetic activity of the K-dwarf component in V471\,Tau}

\author{Zs.~K\H{o}v\'ari\inst{1}
        \and
          L.~Kriskovics\inst{1}\thanks{Bolyai J\'anos Research Fellow}
        \and
          K.~Ol\'ah\inst{1}
        \and
          P.~Odert\inst{2,3}
        \and
         M.~Leitzinger\inst{2,3}
        \and
          B.~Seli\inst{1}
        \and
          K.~Vida\inst{1}
        \and
        T.~Borkovits\inst{1,4}
        \and
        T.~Carroll\inst{5}
  }

   \institute{Konkoly Observatory, Research Centre for Astronomy and Earth Sciences, Konkoly Thege \'ut 15-17., H-1121 Budapest, Hungary\\
        \email{kovari@konkoly.hu}
        \and
           Institute of Physics/IGAM, University of Graz, Universit\"atsplatz 5, A-8010 Graz, Austria
        \and
           Space Research Institute, Austrian Academy of Sciences, Schmiedlstra\ss{}e 6, A-8042 Graz, Austria
        \and
        Baja Astronomical Observatory of University of Szeged, Szegedi \'ut, Kt. 766, H-6500 Baja, Hungary
        \and        
           Leibniz-Institute for Astrophysics (AIP), An der Sternwarte 16, D-14482 Potsdam, Germany
            }

   \date{Received ...; accepted ...}

% \abstract{}{}{}{}{} 
% 5 {} token are mandatory
 
  \abstract
   {Late-type stars in close binary systems can exhibit strong magnetic activity owing to rapid rotation supported by tidal locking. On the other hand, tidal coupling may suppress the differential rotation which is a key ingredient of the magnetic dynamo.}
   {We scrutinize the red dwarf component in the eclipsing binary system V471\,Tau in order to unravel relations between different activity layers from the stellar surface through the chromosphere up to the corona. We aim at studying how the magnetic dynamo in the late-type component is affected by the close white dwarf companion.}
   {We use space photometry, high resolution spectroscopy and X-ray observations from different space instruments to explore the main characteristics of magnetic activity.  We apply a light curve synthesis program to extract the eclipsing binary model and further analyze the residual light variations. Photometric periods are obtained using a Fourier-based period search code. We search for flares by applying an automated flare detection code. Spectral synthesis is used to derive or specify some of the astrophysical parameters. Doppler imaging is used to reconstruct surface temperature maps, which are cross-correlated with each other to derive surface differential rotation. We apply different conversion techniques to make it possible to compare the X-ray emissions obtained from different space instruments.}
   {From \emph{K2} photomery we find that 5-10 per cent of the apparent surface of the red dwarf is covered by cool starspots. From seasonal photometric period changes we estimate a weak differential rotation. From the flare activity we derive a cumulative flare frequency diagram which suggests that frequent flaring could have a significant role in heating the corona. Using high resolution spectroscopy we reconstruct four Doppler images for different epochs which reveal an active longitude, that is, a permanent dominant spot facing the white dwarf. From short term changes in the consecutive Doppler images we derive a weak solar-type surface differential rotation with $\alpha_{\rm DR}=0.0026$ shear coefficient, similar to that provided by photometry. The long-term evolution of X-ray luminosity reveals a possible activity cycle length of $\approx$12.7\,ys, traces of which were discovered also in the H$\alpha$ spectra.}
   {We conclude that the magnetic activity of the red dwarf component in V471\,Tau is strongly influenced by the close white dwarf companion. We confirm the presence of a permanent dominant spot (active longitude) on the red dwarf facing the white dwarf. The  weak differential rotation of the red dwarf is very likely the result of tidal confinement by the companion. We find that the periodic appearance of the inter-binary H$\alpha$ emission from the vicinity of the inner Lagrangian point is correlated with the activity cycle.}


   \maketitle
%
%-------------------------------------------------------------------

\section{Introduction}

Magnetic fields have a strong effect on stellar structure and overall, on the long-term evolution of stars, including our Sun. Studying the manifestations of stellar magnetic activity, from photospheric starspots through the bright chromospheric features to the active corona, helps us understand the nature of the underlying magnetic dynamo, and eventually the magnetic evolution of stars. In the last few years it became feasible to study flares on a large number of stars with the \emph{Kepler}, \emph{K2} and ongoing \emph{TESS} missions \citep[e.g.][]{2015MNRAS.447.2714B}. Such continuous high-precision space photometry combined with high-resolution spectroscopic observations can reveal the connection between the occurrence rate of flares and spot distributions.

It has been learned that late-type stars in close binary systems can exhibit even stronger magnetic activity owing to rapid rotation supported by tidal locking \citep[e.g.][]{2014MNRAS.444..192H}. Tidal coupling, however, may suppress the differential rotation \citep[][]{1982ApJ...253..298S} which is a key ingredient of the magnetic dynamo. In the recent study by \citet[][]{2017AN....338..903K} it was shown that the tidal effect of a close companion star has indeed a suppressive effect on differential rotation.

In our study we scrutinize the red dwarf component in the eclipsing binary system V471\,Tau in order to explore relations between different activity layers from the stellar surface through the chromosphere up to the corona. V471\,Tau appeared first in the literature as a spectroscopic binary in the General Catalogue of Stellar Radial Velocities \citep[GCVR,][]{1953GCRV..C......0W}. The system located in the Hyades star cluster consists of DA white dwarf primary with a K2V companion, forming a post-common envelope binary (PCEB) star.
The age of the Hyades cluster of $t\approx 680\,\mathrm{Myr}$ provides an upper limit for the age of V471\,Tau \citep{2018ApJ...863...67G}.
The binary is dubbed to be a pre-cataclysmic variable as well, which means that there is no significant mass transfer in the system, because the K star does not fill its Roche lobe. The orbital period of the eclipsing system of 0.52118\,d \citep[for updated orbital parameters see][]{2015ApJ...810..157V} is modulated on the long term, which was explained by a light-time effect due to the gravitational influence of a third body, possibly a brown dwarf \citep{2001ApJ...546L..43G}.
As an alternative explanation, the Applegate mechanism was offered by \citet{2015ApJ...800L..24H}, where eclipse timing variations were interpreted as changes in the quadrupole moment within the K2V star.
Recently, \citet{lanza_v471} has proposed a mechanism based on a permanent non-axisymmetric gravitational quadrupole moment. It is no exaggeration to say that V471\,Tau is an actual astrophysical laboratory for studying many aspects of stellar evolution.
The white dwarf primary with a surface temperature of $\approx$35000\,K has strong emission in the ultraviolet (UV), extreme ultraviolet (EUV) and X-ray regimes. In such PCEB systems X-ray eemission can originate from either the white dwarf or the corona of the K star. Moreover, the hot primary can also heat the exposed hemisphere of the secondary. V471\,Tau is a well-known variable at the high energy electromagnetic regime as well, being observed during the \emph{Einstein}  \citep{1983ApJ...267..655Y}, \emph{IUE} \citep{1984AJ.....89.1252G}, \emph{EXOSAT} \citet{1986ApJ...309L..27J}, \emph{ROSAT} \citep{1992MNRAS.255..369B,1998MNRAS.297.1145W} and more recently in the course of \emph{Chandra} \citep{2005ApJ...621.1009G} and \emph{XMM Newton} missions, to mention the most important ones. Accordingly, among other major findings, \citet{1986ApJ...309L..27J} discovered of the pulsation of the white dwarf, while \citep[][]{1992MNRAS.255..369B} confirmed a weak accretion from the K star to the white dwarf via stellar wind. Using $UV$ data from Goddard High Resolution Spectrograph (GHRS) on board the \emph{Hubble Space Telescope}  \citet{2001ApJ...560..919B} reported Coronal Mass Ejections (CMEs) from the active K dwarf and estimated a 100-500\,CMEs per day emission rate, i.e. about a hundred times higher frequency compared to the Sun.

As for the surface magnetic activity of the secondary, the K2V star, forced to rotate synchronously with the orbital period, is known to exhibit rotational variability due to starspots \citep[e.g.][]{1986Ap&SS.120...97E} and flare activity \citep{1983ApJ...267..655Y}.
The first Doppler images of the star were presented by \citet{1995AJ....110.1364R}, who obtained four separate temperature maps spanning over a year.
In another Doppler-imaging study \citet{2006MNRAS.367.1699H} found that despite the tidally inhibited differential rotation the K-dwarf component of the V471\,Tau binary system showed indeed a high level of magnetic activity. In our paper we perform a complex study of the magnetic activity of V471\,Tau, including new Doppler images and flare statistics for the spotted red dwarf.
%This was confirmed by spectroscopic observations: the binary shows strong emission in $H\alpha$ and Ca II H\&K lines (ref???)  during eclipses when the active regions on the K star pointing at the white dwarf is facing the observer.

The paper is organized as follows. In Sect.\,\ref{sect:data} we summarize the available photometric and spectroscopic data to be analyzed. Using \emph{Kepler K2} observations, in Sect.\,\ref{sect:K2} by extracting orbital solution we analyze light variations due to spots and flares. In Sect.\,\ref{sect:params} by the means of spectral synthesis we derive precise astrophysical parameters of the active K-dwarf component, which is used to perform a Dopper imaging study in Sect.\,\ref{sect:di} where we estimate the surface differential rotation of the spotted star. The chromospheric and coronal activity of the K star is analyzed in Sect.\,\ref{sect:xray}. Our results are discussed in Sect.\,\ref{sect:disc} and summarized in Sect.\,\ref{sect:summary}.

\section{Data}
\label{sect:data}
\subsection{K2 photometry}

\begin{figure}[th]
    \includegraphics[width=\columnwidth]{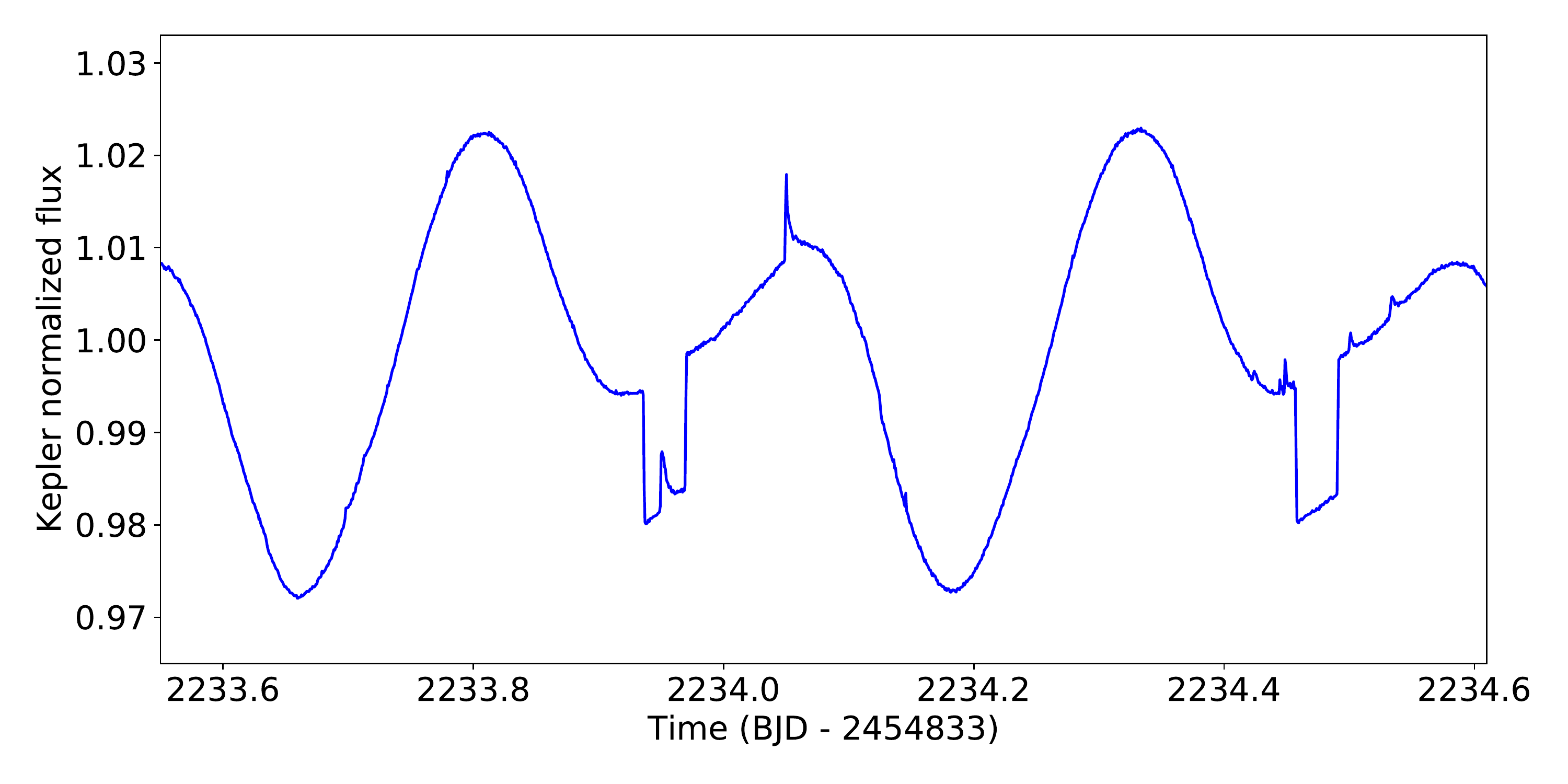}
      \caption{Part of the \emph{Kepler K2} short cadence data of V471\,Tau with a particular flare event occured when the white dwarf component was in eclipse.
      }
        \label{fig:k2scflare}
\end{figure}

The \emph{Kepler K2} mission provided long-term high-precision space photometry  \citep{2014PASP..126..398H} of V471\,Tau (EPIC\,210619926). The star was observed as part of the 4th Campaign of the mission between 08 February$-$20 April 2015 (BJD\,2457061.7910$-$2457132.6877), covering 136 subsequent orbital cycles.
Both short (1-min) and long cadence (30-min) light curves of V471\,Tau are available for the interval. Although the \emph{K2} data do not coincide with the spectroscopic observations used in this paper, the short cadence space data are suitable for studying spot evolution and flare activity on a somewhat longer timescale. As an example, in Fig.\,\ref{fig:k2scflare} we plot a two cycles long part of the short cadence light curve, where in-eclipse and out-of-eclipse flares are also seen.

\subsection{High-resolution spectroscopy}
We use high-resolution ($R$=81000) spectroscopic observations of V471\,Tau publicly available in the data archive of the Echelle SpectroPolarimetric Device
for the Observation of Stars (ESPaDOnS) at Canada--France--Hawaii Telescope (CFHT). The spectra cover the 3700--10500\,\AA\ optical range.
The available data were retrieved in two seasons, the first in December 2005 and the second at the turn 2014/2015. We selected 297 %301-4
and 222 %228-6
spectra from the two seasons, respectively, to achieve uttermost subsets (in the long run two in both seasons) with regard to optimal phase coverage for Doppler imaging (see Sect.\,\ref{sect:di}). Low quality spectra with signal-to-noise S/N$<$40 were omitted from the selection.
Before using the four data subsets for Doppler image reconstructions  we shifted the spectra to the rest wavelength and continuum fitting was carried out. The CFHT spectra contain the H$\alpha$ region which is used to study the chromospheric activity and discover possible mass motions in the binary (see Sec.\,\ref{sect:halpha}).
The observing log of the CFHT-ESPaDOnS  spectra used in our study is given in Table\,\ref{table:cfht} in Appendix\,\ref{app1}.
%low quality data selection  with signal-to-noise ratios S/N$\geq$50 (i.e., ).  with  The first observing
%HJD=2453718.714 and 2453723.028 14$-$20  and the second between 20 December 2014$-$27 February 2015
%consist of 336 spectra .
%Data from the second observing run between HJD=2457xxx.xxx and 2457xxx.xxx  consist of xxx good quality (S/N$\geq$50) spectra. The typical signal-to-noise ratio in the second run was S/N$\approx$80??. Anew, low quality spectra (S/N$\leq$50) were left out.

\subsection{X-ray observations}
The V471\,Tau system has been detected with many X-ray instruments during the last decades. Here we attempt to infer the long-term variability of the K dwarf's coronal emission. As none of the X-ray observations coincide with the epochs of the high resolution spectroscopic data for Doppler imaging, we do not address its variability on shorter timescales like rotational modulation due to the distribution of active regions. We note that the binary components are not spatially resolved in the X-ray observations, but it is well established that only the softer ($>50$\AA) wavelength range is affected by the emission from the white dwarf \citep{2003ApJ...594L..55D,2005ApJ...621.1009G}, therefore its eclipse (at $\phi$=0 orbital phase) is only observed in soft X-ray/EUV light curves \citep{1996aeu..conf..349C,1998MNRAS.297.1145W}. What makes the investigation of the long-term coronal variability difficult is that V471\,Tau was rarely observed by the same instrument more than once, the only exception being the \emph{ROSAT} pointed survey (twice during the same year) and \emph{XMM Newton} (separated by 15 years). Moreover, the K dwarf is very active due to its high rotation rate, so its emission is variable also on short time scales, including frequent flares, which makes the definition of a quasi-quiescent emission level during a given epoch difficult.

For the analysis we select data from five X-ray instruments covering the years 1991-2019 (Table\,\ref{xrayobs}). We do not include data from earlier instruments (\emph{Exosat}, \emph{Einstein}) and the \emph{ROSAT} All Sky Survey (RASS), because of their softer wavelength coverage (including mainly the white dwarf's emission) and/or their short duration (RASS). 
When selecting the space data we considered the epochs of our new Doppler reconstructions as well (see Sect.\,\ref{sect:di}).
In Sect.\,\ref{sect:xray} we describe the detailed analysis steps for each instrument and observation, as well as which data were taken from previous analyses in the literature.

\begin{table*}[thb!!]
\caption{Adopted X-ray observations of V471\,Tau with their durations and the quiescent periods within them}
\label{xrayobs}
\begin{tabular}{ccccrr}
\hline\noalign{\smallskip}
Date [dd.mm.yyyy] & Mission & Instrument & Observing ID & Duration [s] & Quiet [s] \\
\hline\hline\noalign{\smallskip}
23.08.1991 & \emph{ROSAT} & PSPC & RP200107M01 & 27692 & all \\
26.08.1996 & \emph{ASCA} & SIS+GIS & 24032000 & 102416 & all \\
14.08.1997 & \emph{ROSAT} & HRI & RH202309N00 & 15793 & 1104 \\
24.01.2002 & \emph{Chandra} & HRC-S/LETG & 2523 & 87990 & all \\
01.08.2004 & \emph{XMM Newton} & EPIC/PN & 0203260101 & 60924 & 18783 \\
30.07.2015 & \emph{Swift} & XRT & 33909 & 2447 & all \\
04.09.2019 & \emph{XMM Newton} & EPIC/PN & 0844350101 & 63500 & 9381 \\
\hline
\end{tabular}
\end{table*}

\section{Stellar properties}\label{sect:params}

\subsection{Adopted parameters}
The \emph{Gaia} DR2 parallax of $\pi=20.957\pm0.044\,\mathrm{mas}$ \citep{gaia_dr2}
yields a distance of $d=47.72\pm0.10\,\mathrm{pc}$ for V471\,Tau.
This distance 
assuming $V_{\rm br}\approx9\fm35$
%%9.35+0.02 V-ben a wd jarulekaval csokkentve
for the brightest $V$ magnitude observed so far in 2004 \citep[cf.][]{2005MNRAS.360.1077I} and removing $\Delta V_{\rm wd}=-0\fm02$ as the contribution of the white dwarf \citep[][]{1981AcA....31...37R} and taking into account a maximum of $A_V=0\fm003$ insterstellar exctinction  \citep[cf.][]{2006AJ....132.2453T}
yields an absolute visual magnitude $M_V=5\fm97\pm0\fm03$. Taking $BC=-0.257$ bolometric correction from \citep{flower_bc} gives a bolometric magnitude $M_{\mathrm{bol}}=5\fm72 \pm0\fm03$. This gives $L/L_{\odot}=0.41\pm0.01$ when using $M_{\mathrm{bol},\odot}=4\fm74$ value for the Sun. On the other hand, when adopting $R/R_{\odot}=0.91\pm0.02$  \citep[][see their Table 11]{2015ApJ...810..157V} and with $T_{\mathrm{eff}}=4980$\,K (see Sect.\,\ref{sme}) the Stefan--Boltzmann law would give a bit higher $L/L_{\odot}$ ratio of 0.46$\pm$0.04; still
in a fair agreement with the value obtained from the bolometric magnitude.

\begin{figure}[thb]
    \includegraphics[width=\columnwidth]{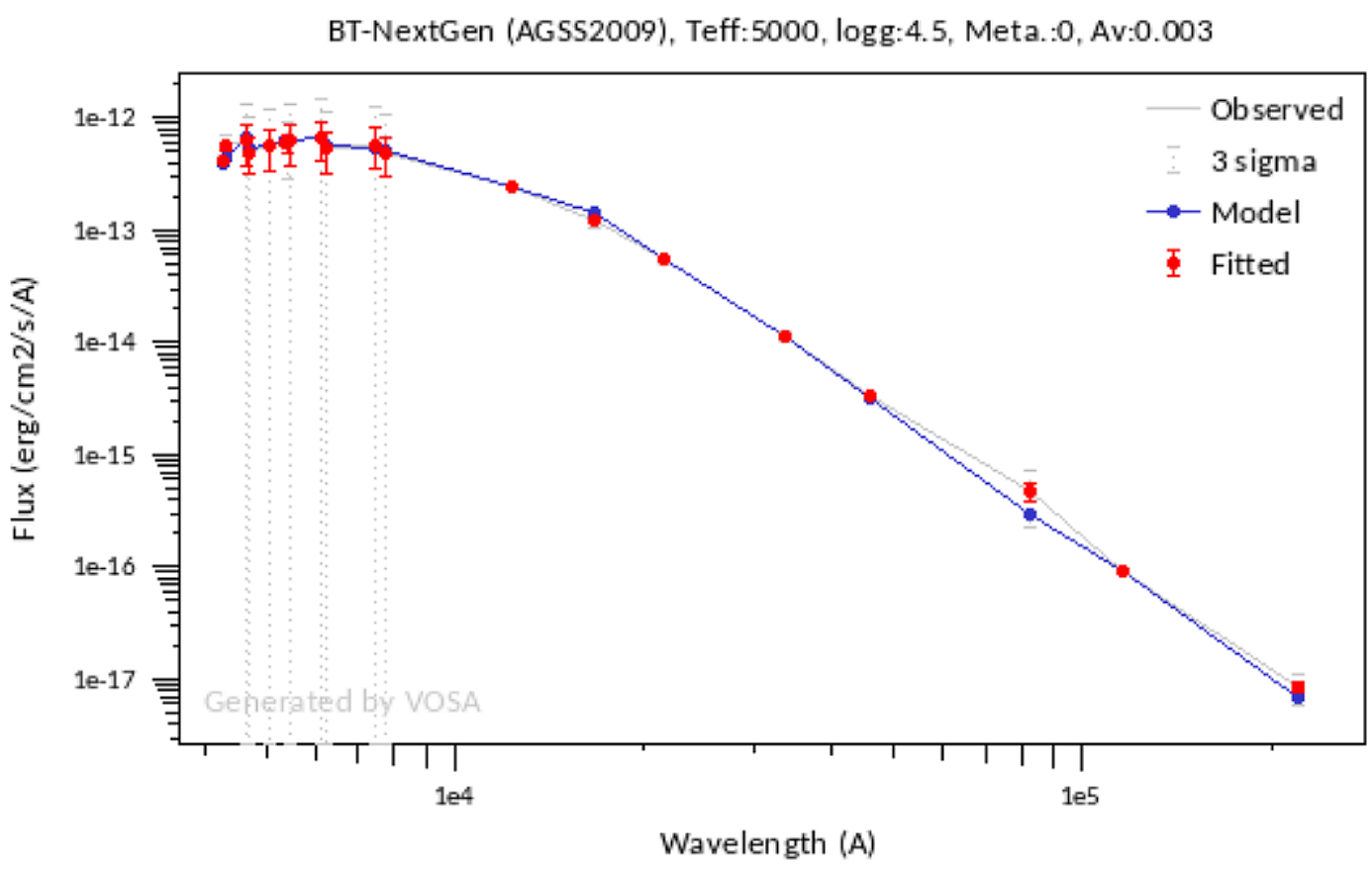}
      \caption{Spectral energy distribution for V471\,Tau generated by the VOSA SED analyzer tool \citep{2008A&A...492..277B}. The synthetic spectrum (blue line) is fitted to the available archival photometry (red dots) from \emph{Tycho}, \emph{SLOAN/SDSS}, \emph{Gaia} DR2, \emph{2MASS}, \emph{AKARI/IRC}, and \emph{WISE} surveys.
      }
        \label{fig:sed}
\end{figure}

We perform a spectral energy distribution (SED) synthesis using the virtual observatory (VO) tool VOSA \citep{2008A&A...492..277B} to build the SED of V471\,Tau from the available VO catalogs. We use VOSA to collect archival photometry from \emph{Tycho}, \emph{SLOAN/SDSS}, \emph{Gaia} DR2, \emph{2MASS}, \emph{AKARI/IRC}, and \emph{WISE} surveys. As input we confined the effective temperature to 4950\,K<$T_{\rm eff}$<5050\,K, the surface gravity to 4.25<$\log g$<4.75, and the metallicity to $-$0.2<$[\mathrm{Fe}/\mathrm{H}]$<0.2 (cf. Sect.\,\ref{sme}). The resulting synthetic spectrum plotted in Fig.\,\ref{fig:sed}  corresponds to $L/L_{\odot}$ of 0.40$\pm$0.21, in agreement with both values given hereinabove.

The $v\sin{i}=91$\,$\mathrm{km\,s}^{-1}$ projected rotational velocity of the K2 star was adopted from \citet[][see also their references]{2006MNRAS.367.1699H}. The  inclination $i=77\fdg4^{+7\fdg5}_{-4\fdg5}$ of the binary orbit with respect to the line of sight was derived in \citet{2001ApJ...563..971O}, which we adopt here for the K-dwarf with the assumption that its rotational axis is perpendicular to the orbital plane.

Adopting $P_{\rm orb}=P_{\rm rot}$ period and HJD$_0=2440610.06406$ reference time (mid-primary minimum, when the white dwarf is in eclipse) from \citet{2001ApJ...546L..43G} we use the following equation for calculating the phase values of the spectroscopic observations in Table\,\ref{table:cfht}:
\begin{equation}\label{eq:phase}
    \mathrm{HJD}_{\phi=0}=2440610.06406+0\fd521183398\times E.
\end{equation}

\subsection{Spectral synthesis}\label{sme}

We carried out a detailed spectroscopic study based on spectral synthesis using the code SME \citep{piskunov_sme}. During the synthesis, MARCS models were used \citep{gustafsson_marcs} and atomic paramaters were taken from VALD \citep{kupka_vald}. Macroturbulence was computed using the following equation from \citet{valenti_macro}:
\begin{equation}\label{eq_vmac}
    v_{\mathrm{mac}}\,\ [\mathrm{km\,s}^{-1}] =3.98-\frac{T_{\mathrm{eff}}\,\ [K]-5770}{650}.
\end{equation}

First, we selected 15 good quality spectra observed in eclipses, when only the K2 star was seen, and performed the synthesis independently for the 15 spectra to get $T_{\mathrm{eff}}$, $\log{g}$, and $[\mathrm{Fe}/\mathrm{H}]$. We used the initial values as input for subsequent iterations. We gradually confined the microturbulence and the projected rotational velocity at $v_{\mathrm{mic}}=0.7 (\pm0.2)$\,$\mathrm{km\,s}^{-1}$ and $v\sin{i}=91 (\pm1)$\,$\mathrm{km\,s}^{-1}$, respectively, in agreement also with literary values \citep[cf., e.g.,][and their references]{2006MNRAS.367.1699H,2012A&A...547A..36A}. Finally, the surface gravity was kept fixed at $\log{g}=4.5$, in agreement with former, well established values \citep[see][and their references]{2015ApJ...810..157V}.
At the end, the procedure resulted in
$T_{\rm eff}=4980$\,K and
$[\mathrm{Fe}/\mathrm{H}]=0.12$. For details of the iterative fitting method see \citet{kriskovics_v1358ori}. The final astrophysical parameters with their errors are listed in Table\,\ref{table_pars}.

\begin{figure}[bht]
    \includegraphics[width=\columnwidth]{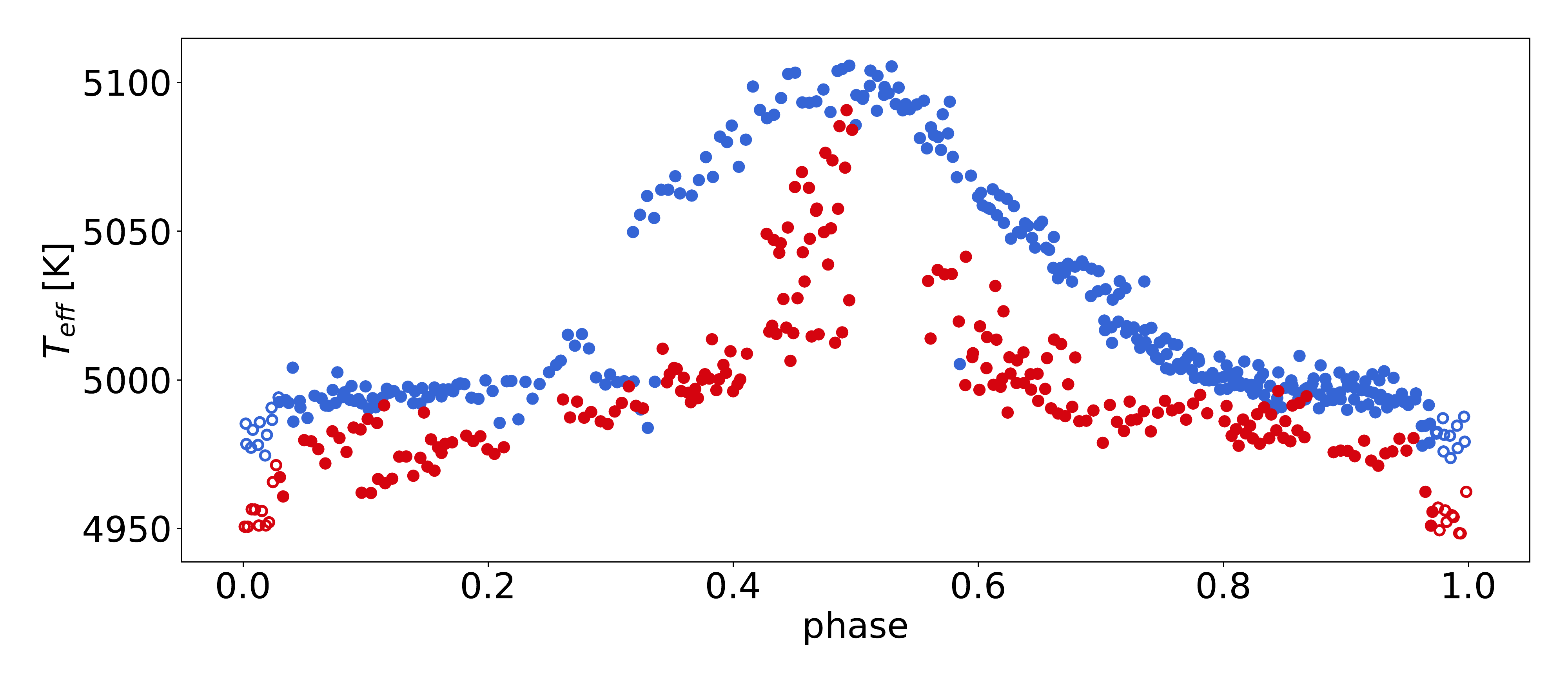}
      \caption{Variation of the surface temperature of the K2 star along the orbital phase. Dots show the temperature values  derived individually for each spectra (blue for the 2005 season, red for the 2014/15 spectra) by spectral synthesis.
      Open circles around zero phase correspond to those unaffected spectra that were observed when the white dwarf was eclipsed. The temperature rise of $\approx$100\,K at $\phi\approx0.5$ in both seasons may indicate the irradiation effect of the hot component. See Sect.\,\ref{sme} for details.
      }
        \label{fig:teffvar}
\end{figure}

In the second test we run the synthesis for all the available CFHT spectra to see if the surface temperature varies along the orbital phase. Therefore, all of the previously derived parameters were kept fixed except for the surface temperature. The results  are plotted in Fig.\,\ref{fig:teffvar}, demonstrating that the surface of the red dwarf is indeed the least affected around $\phi=0.0$ when the hot component is obscured. The temperature rise of $\approx$100\,K at $\phi=0.5$ compared to the eclipse either in 2005 (blue dots in the figure) or 2014/15 (red dots) may be due to the irradiation of the hot component, but surface activity can also play a role (see Sect.\,\ref{Teffrise} for a discussion). Indeed, it is very likely that the stellar dynamo in the synchronously rotating red dwarf is affected by the close companion and therefore manifestations of magnetic activity may be coupled to the orbital phase \citep[cf.][]{1991A&A...251..183S}.

\begin{table}[thb!!]
%\begin{center}
\caption{Fundamental astrophysical data of the K2 dwarf component of V471\,Tau}
\label{table_pars}
%\centering 
 \begin{tabular}{ll}
  \hline\noalign{\smallskip}
  Parameter               &  Value \\
  \hline\hline\noalign{\smallskip}
Spectral type    & K2\,V  \\
Gaia distance [pc]  &  $47.72\pm0.10$\\
$V_{\rm br}$ [mag]  & $9\fm37\pm0\fm03$ \\
%$(B-V)_{\rm HIP}$  [mag] & $ ?.??8$ \\
$M_{\rm bol}$  [mag] & $5.72\pm0.03$ \\
Luminosity [${L_{\odot}}$] & $0.41\pm0.01$ \\
$T_{\rm eff}$ [K]  & $4980\pm10$\\
$\log{g}$ (in cgs)  & $4.50\pm0.05$ \\
$[\mathrm{Fe}/\mathrm{H}]$& $0.12\pm0.04$ \\
$v_{\mathrm{mic}}$ [$\mathrm{km\,s}^{-1}$] & $0.7\pm0.2$ \\
$v_{\mathrm{mac}}$\tablefootmark{a} [$\mathrm{km\,s}^{-1}$] & $ 5.2$\,\\
$v\sin{i}$ [$\mathrm{km\,s}^{-1}$] & $91\pm1$\\
Rotation period [d] & $0.521183398 $ \\
Inclination\tablefootmark{b} [\degr] &   $77.5^{+7.5}_{-4.5}$ \\
Radius\tablefootmark{c} [$R_{\odot}$] & $0.91\pm0.02$   \\
Mass\tablefootmark{c} [$M_{\odot}$] & $0.95\pm 0.05$   \\
\hline
\end{tabular}
\tablefoot{
\tablefoottext{a}{Computed using Eq.\,\ref{eq_vmac};}
\tablefoottext{b}{\citet{2001ApJ...563..971O};}
\tablefoottext{c}{\citet{2015ApJ...810..157V}.}
}
%\end{center}
\end{table}

\section{Photometric analysis}\label{sect:K2}
\subsection{Extracting the eclipsing binary light curve}\label{sect:lcfactory}

\begin{figure*}[th]
    \includegraphics[width=\textwidth]{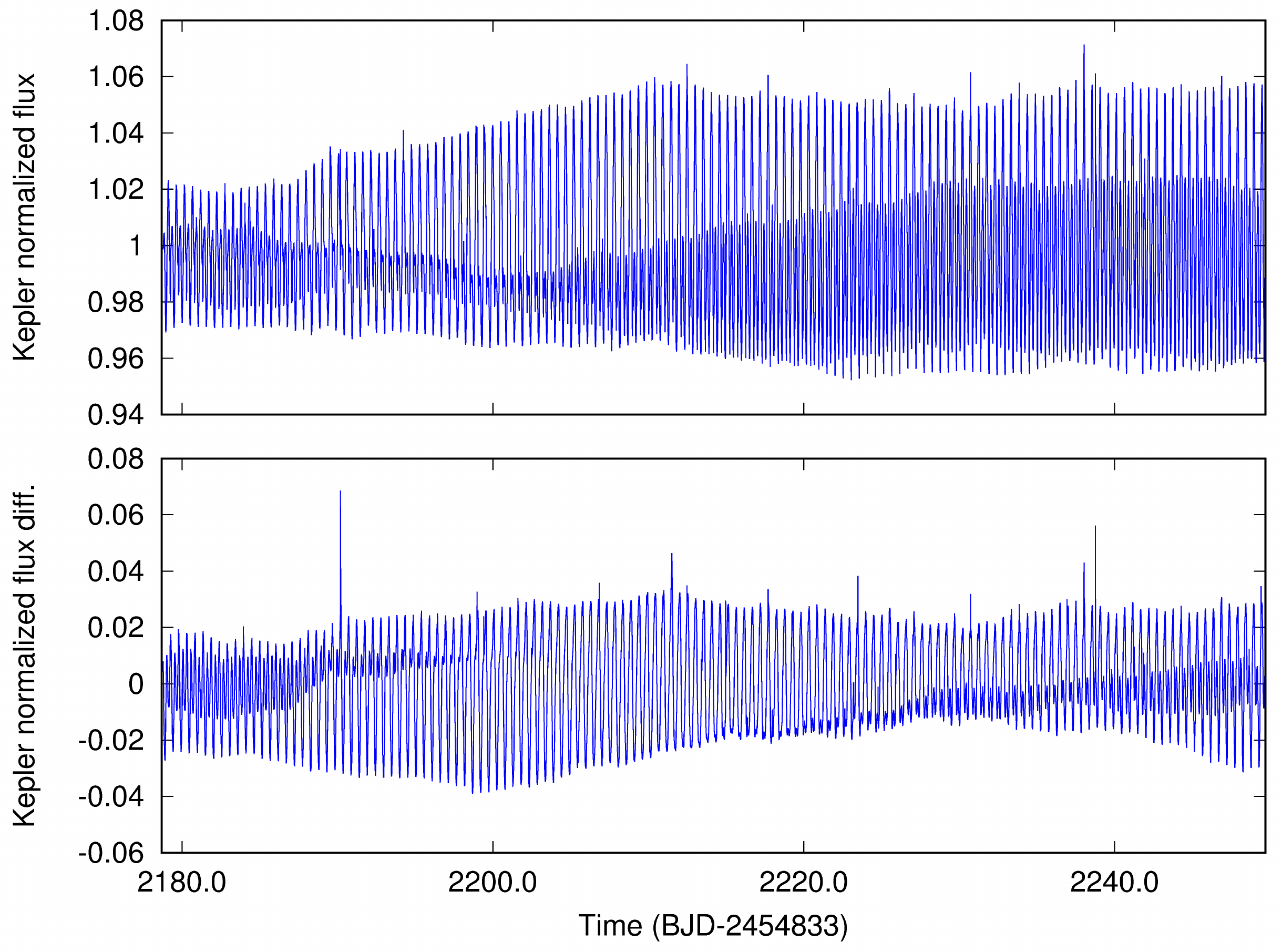}
      \caption{\emph{K2} light curve of V471\,Tau (top) and residual light variations (bottom) due to spots and flares after extracting the eclipsing binary model. See Sect.\,\ref{sect:lcfactory} for details.
      }
        \label{fig:lcfactory}
\end{figure*}

Before analyzing the photometric signs of the magnetic activity of V471\,Tau we removed the light curve variations caused by the close binary nature as follows. First, we formed a phase-folded light curve with the use of the \emph{K2} short cadence light curve. The data were grouped into 2000 orbital phase bins of equal duration. Then, the average of each phase cells were calculated and rendered to the mid-phase point of the given cell. 
This way the fastest part of the non-orbital phase-locked light variations were removed, but a significant amount of the intrinsic variability over longer timescales has remained in the light curve.
%This phase-folded, binned and averaged light curve though averaged out the fastest part of the non orbital phase locked light variations, but a significant amount of the intrinsic variations varying slower in time has remained in the light curve. 
In order to remove these slower variations we synthesized a pure binary light curve with the use of the software package {\tt Lightcurvefactory} \citep[][]{2020MNRAS.493.5005B}. The input parameters were taken from \citet{2015ApJ...810..157V}. Besides the eclipses, the binary model has included ellipsoidal variations and the effect of Doppler-boosting, too. We did not consider, however, the reflection/irradiation effect of which the photometric contribution was found to be negligible.
Using the parameters from \citet[][see their Table 8, 4th column]{2015ApJ...810..157V} after a natural readjustment of the mid-time of a given eclipse ($T_0$) parameter to the value appropriate for \emph{K2} data and comparing the synthesized light curve against the folded, binned, averaged \emph{K2} light curve, the residual curve has shown significant shoulders around the partial phases of the eclipses, revealing, that the durations of both the total eclipses and the totality should slightly be readjusted. We made it manually with slight modifications of the radius of the K dwarf component. This way we were able to remove the shoulders from the residual file. Finally, with a repeated use of {\tt Lightcurvefactory}, this slightly refined synthetic light curve was removed from the original, $\approx$71\,d long \textit{K2} measurements.
The resulting residual light curve is plotted in Fig.\,\ref{fig:lcfactory}.

\subsection{Analyzing spot variability}\label{sect_k2diffrot}

According to Fig.\,\ref{fig:lcfactory} the light curve is solidly modulated by a continuously changing spot distribution. The relative intensity of the spot amplitude varies between 0.03-0.06. From this we estimate that 5-10 per cent of the apparent stellar surface is covered by spots \citep[see Eq.\,3 in][]{2019ApJ...876...58N}. Assuming surface differential rotation, the change in the rotational frequency signals can be interpreted as a change in the latitude of the dominant surface spots. To recover the presence of differential rotation from the 136-cycle ($\approx$71-day) long photometric time series we apply the following technique. First, the entire residual light curve (bottom panel of Fig.\,\ref{fig:lcfactory}) was divided into eight consecutive sections of $\approx$9-day long (17 rotational cycles), i.e., a compromise to avoid averaging frequency signals over a longer time scale but still get stable/reliable frequencies for the given part. To derive the most characteristic rotational period for each part we apply the Fourier-based period search code MuFrAn \citep{2004ESASP.559..396C}. The resulting short term rotational periods listed in Table\,\ref{tab_K2periods} indicate that the change in the period remains within a narrow interval of $\Delta P\approx0.001$\,day. Dividing this by the average rotation period from Table\,\ref{table_pars}, we can estimate the dimensionless surface shear coefficient as  $|\alpha_{\mathrm{DR}}|\gtrsim\Delta P_{\rm phot}/\overline{P}$, where $\Delta P_{\rm phot}$ is the
full range of the seasonal photometric period associated with surface spots, while $\overline{P}$ is the average period over long term. From the values listed in Table~\ref{tab_K2periods},
we obtain $|\alpha_{\mathrm{DR}}|\gtrsim0.002$, i.e, a weak shear, almost solid-body rotation. We note, that this method alone does not allow to determine the sign of the surface shear parameter, i.e., whether the differential rotation is solar type or antisolar.

% ---------------------------- T1
\begin{table}
% \centering%%%
\caption{Seasonal photometric periods from $K2$ data}
\label{tab_K2periods}
%\begin{footnotesize}
\begin{tabular}{c c c c}
\hline\noalign{\smallskip}
Season & BJD & BJD & $P_{\rm phot}$ \\
No. & start\tablefootmark{a} & end\tablefootmark{a} & [d] \\
\hline\hline
\noalign{\smallskip}
1 & 2228.790988 & 2236.999957 &  0.520938 \\
2 & 2237.000638 & 2245.999560 &  0.520565 \\
3 & 2246.000241 & 2254.999765 &  0.520193 \\
4 & 2255.000446 & 2263.999913 &  0.520645 \\
5 & 2264.000594 & 2272.999345 &  0.520336 \\
6 & 2273.000026 & 2281.999446 &  0.519983 \\
7 & 2282.000127 & 2290.999555 &  0.520641 \\
8 & 2291.000236 & 2299.687090 &  0.519987 \\
\hline
\end{tabular}
\tablefoot{
\tablefoottext{a}{+2454833.0}
%\tablefoottext{b}{$\sigma_{P_{\rm phot}}=0.008$.}
}
\end{table}

\subsection{Finding flares}\label{sect:flares}

After extracting the eclipsing binary model from the \emph{K2} light curve we search for flares in residual time series (bottom panel of Fig.\,\ref{sect:lcfactory}) by applying the automated flare detection code FLATW’RM \citep[FLAre  deTection With Ransac Method,][]{2018A&A...616A.163V}. The code enables the user to adjust the detection level and the number of successive points associated with a given flare. This is especially important when detecting the smallest events just above the noise level, which are always difficult to find. However, 
%{\vidak{even after removing the eclipsing binary model from the \emph{K2} 
%light curve, a systematic, short-period variation remained, which cannot be described with polynomials of reasonable orders. This increases the variation -- "noise" -- of the light curves.}}
close to the noise limit the number of false positives is starting to increase, which, on the other hand, can be compensated by increasing the number of associated points. After a visual inspection, however, we declared 15 false positives among the detected events. On the other hand, a similar number of possible flares just above the noise level could have been missed. Due to these reasons, we ran a second analysis using a flare-searching neural network (Vida et al., in prep.), and corrected its output manually for possible false positive/negative detections. This analysis yielded a final number of 198 confirmed flare events in the 71-day long \emph{K2} time series.

\begin{figure}%[bht]
    \includegraphics[width=\columnwidth]{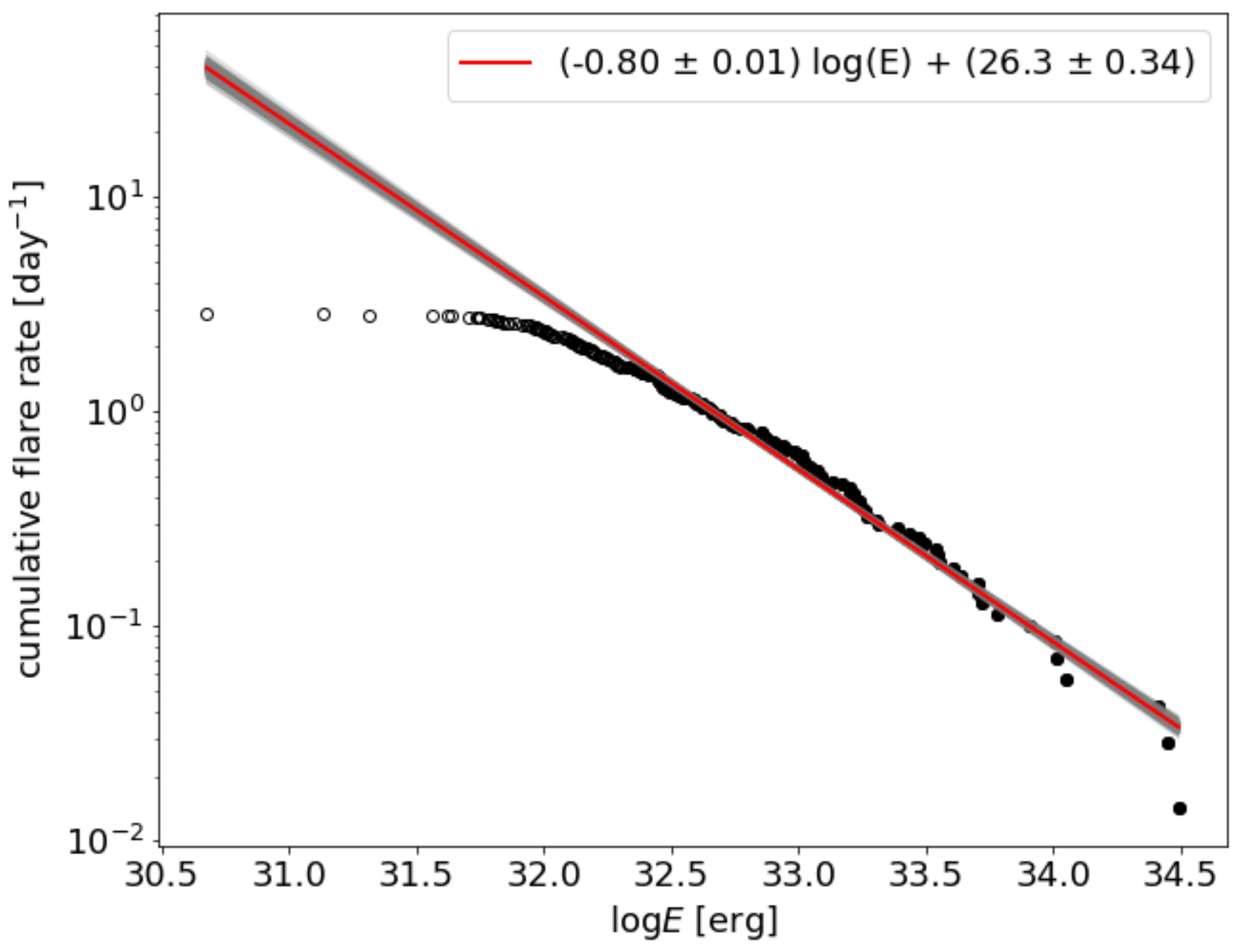}
      \caption{Cumulative flare-frequency  diagram for V471\,Tau. The slope of the linear fit (orange line) corresponds to $\alpha=1.8$ power law index.
      }
        \label{fig_ffd}
\end{figure}

Taking $T_{\rm eff}$ and $R$ from Table\,\ref{table_pars} the blackbody approximation yields $L_{\rm q}=4.64\times10^{32}$\,erg\,s$^{-1}$ quiescent luminosity for the K dwarf through the \emph{Kepler} filter. To determine the individual flare energies, first, the background variation (due to spots) is removed by fitting a low-order polynomial to the 1 hour interval of the given flare (omitting the flaring points). The order of the polynomial is determined by the Bayesian Information Criterion. After subtracting this polynomial, the net flare light curve is integrated over the duration of the flare, yielding the $\varepsilon_{\rm f}$ relative flare energy (or equivalent duration) of the event. Multiplying $\varepsilon_{\rm f}$ with $L_{\rm q}$ gives the total energy $E_{\rm f}$ released by the flare. Finally, $E_{\rm f}$ values are used to make up the cumulative flare frequency diagram in Fig.\,\ref{fig_ffd}. 
%To determine the flare energies, first, the background variation is removed by fitting a polynomial to the Ransac-inlier points (that will not include the flare itself). After subtracting this polynomial, the light curve is integrated over the time of the flare, yielding the equivalent duration of the event, which, after multiplying with the quiescent stellar energy, gives the total energy released by the star during the flare.

In principle, the cumulative number of flares vs. flare energy can be described by a power law \citep[][]{1972Ap&SS..19...75G}.
Accordingly, the logarithm of the $\nu(E)$ cumulative number of flares with energy values larger than or equal to $E$ can be written as:
\begin{equation}
\log \nu(E) = c_1 + c_2\log E,
\end{equation}
where $c_1$ is the intercept while $c_2$ is the slope of a linear function. Rewriting the slope $c_2=1-\alpha$ one can introduce $\alpha$, that is the power law index of the flare energy distribution \citep{2014ApJ...797..121H}.
The low energy turnover of the diagram below $\log E\approx32.6$\,[erg] is most probably the result of the detection threshold. %\citep[see][]{}. 
Fitting the distribution above this turnover (see the red line in Fig.\,\ref{fig_ffd}) yields a power law index of $\alpha=1.80\pm 0.01$.
For further discussion about the flare statistics see Sect.\,\ref{sect:disc:flares}.
%quiescent flux: 2.46812e+33 69.759416 [-0.846255 28.076472] Errors [0.016347 0.549522]

\section{Doppler imaging}
\label{sect:di}

\subsection{The state-of-the art imaging code \it{iMap}}
To reconstruct the surface temperature distribution map of V471\,Tau we use the Doppler imaging code \emph{iMap} \citep{2012A&A...548A..95C}. This state-of-the-art code carries out multi-line inversion on a list of photospheric lines between 5000-6750\,\AA. We selected 24 non-blended absorption lines with suitable line-depth, temperature sensitivity and well defined continuum. The stellar surface is modelled on a  $5^{\circ}\times 5^{\circ}$ resolution grid. Each local line profile is computed with a full radiative solver \citep{carroll_solver}. The local line profiles are disk integrated, and the individually modelled,  disk-integrated lines are averaged. Atomic line data are taken from the Vienna Atomic Line Database \citep[VALD][]{kupka_vald}. Model atmospheres are taken from \cite{castelli_mod} and are interpolated for the necessary temperature, gravity or metallicity values. When solving the radiative transfer local thermodynamical equlibrium (LTE) is assumed because spherical model atmospheres would have required too much computational capacity. Neverheless, imperfections due to the LTE approximation in the fitted line shapes are well compensated by the multi-line approach. Additional input parameters are micro- and macroturbulence, and the projected equatorial velocity (cf. Sect.\,\ref{sect:params}).

For the  surface reconstructions \emph{iMap} uses an iterative regularization based on a Landweber algorithm \citep{2012A&A...548A..95C}. 
The iterative regularization has been proven to converge always on the same image solution \citep[for the tests see Appendix A in][]{2012A&A...548A..95C}. Therefore, no additional constraints are imposed for the image reconstruction.

\subsection{Surface temperature maps}
\label{subsect:di}
In order to achieve good enough phase coverage for the $\approx$0.52 day rotational period, we decided to form two subsets for both observing seasons. The first two sets for the 2005 season, henceforth S1 and S2, consist of 149 and 172 spectra, respectively, both covering rougly four rotations ($\approx$2 days). The S3 and S4 subsets for the 2014/15 season consist of 131 and 97 spectra, respectively, however, due to unfavourable data distribution in this season the S3 subset covers $\approx$20 consecutive rotations ($\approx$10 days) while S4 covers ten rotations (5 days); see Table\,\ref{table:cfht} in the Appendix. All four subsets are used to reconstruct one Doppler image. The mean HJDs for the four subsets are 2453719.705, 2453722.211, 2457016.565, and 2457031.375, respectively. The four Doppler images (referred hereafter as S1, S2, S3 and S4) presented below, can be regarded as "snapshots" of the stellar surface at these times.

%S1 mean hjd: 3719.7048 130  163-33
%S2 mean hjd: 3722.2110 171  173-2

%S3 mean hjd: 7016.5647 131
%S4 mean hjd: 7031.3746 97

\begin{figure*}%[]
            
            \vspace{2.1cm}
            \Large{S1}
            \vspace{-2.7cm}
            
            \hspace{0.5cm}\includegraphics[width=0.98\textwidth]{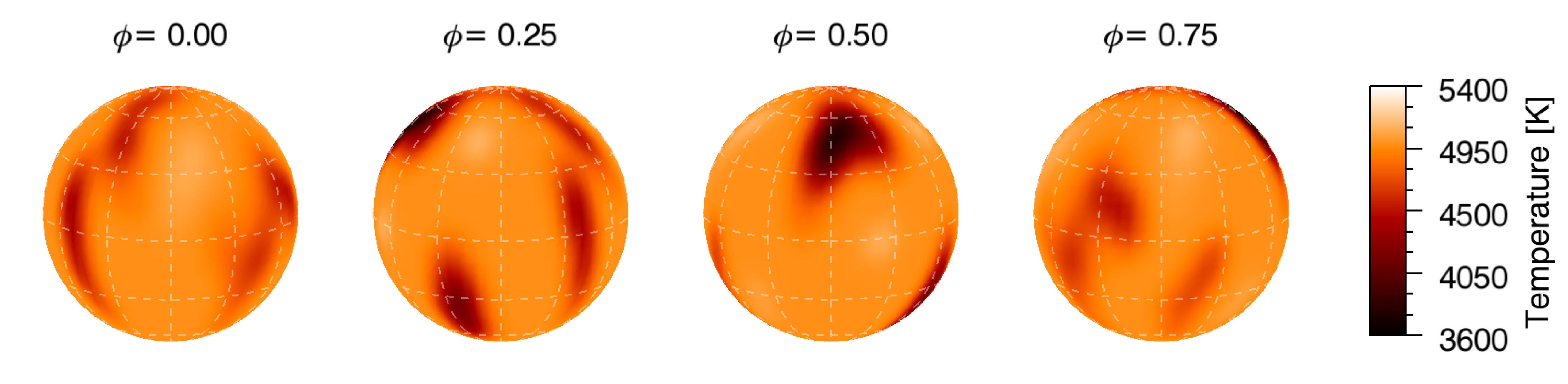}
    
            \vspace{2.1cm}
            \Large{S2}
            \vspace{-2.7cm}
            
            \hspace{0.5cm}\includegraphics[width=0.98\textwidth]{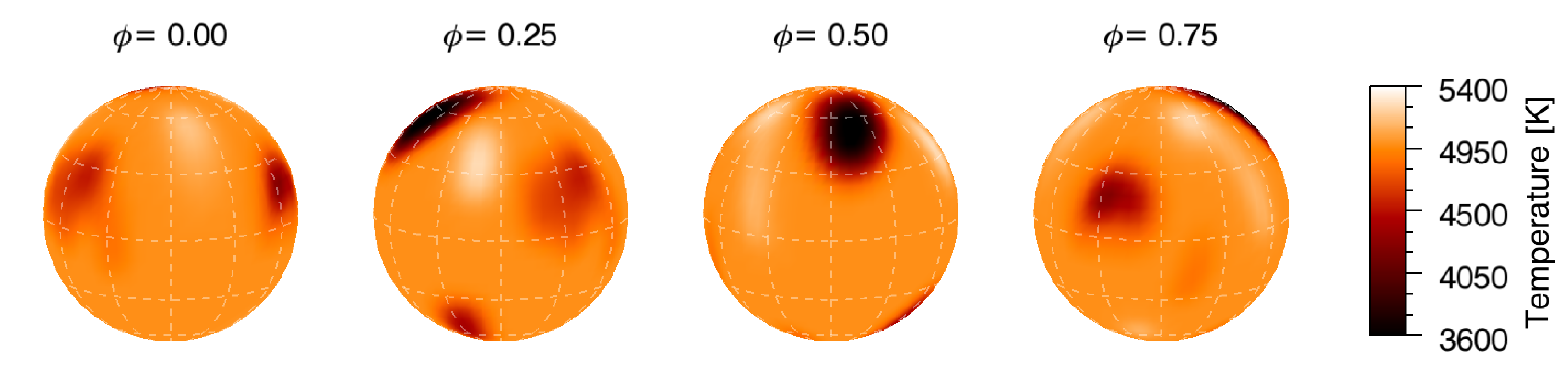}
            
      \caption{Two Doppler images of V471\,Tau from the 2005 data (S1 and S2). The surface temperature reconstructions are plotted in four rotational phases.
              }
         \label{fig_di12}
\end{figure*}
\begin{figure*}%[]
%    \resizebox{\hsize}{!}
             
            \vspace{2.1cm}
            \Large{S1}
            
            \hspace{0.1cm}\Large{+} 
            
            \Large{S2}            
            
            \vspace{-3.2cm}
            \hspace{0.5cm}\includegraphics[width=0.98\textwidth]{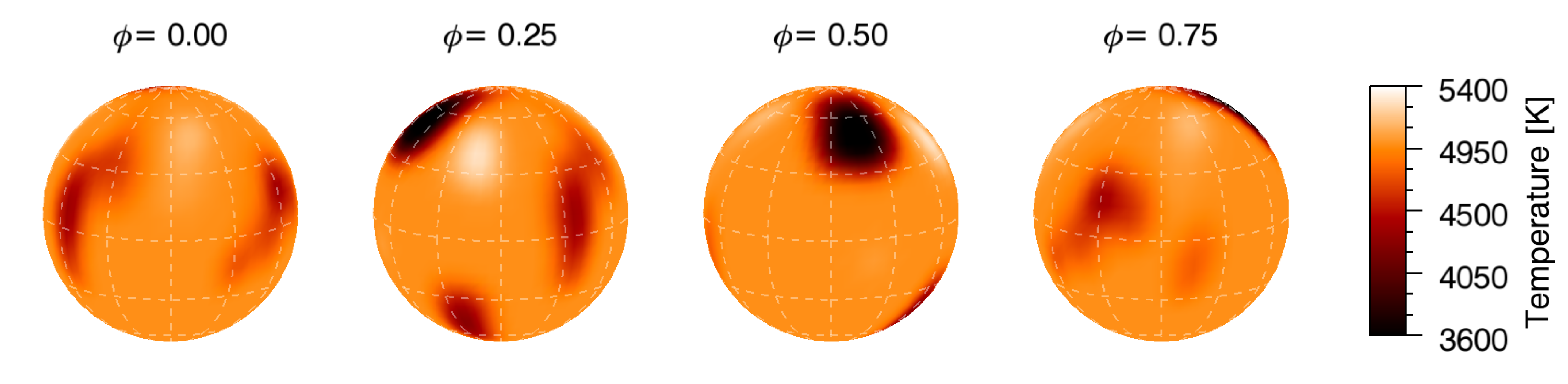}
      \caption{Combined Doppler image of V471\,Tau derived from all the available spectra from 2005 (S1+S2). The surface temperature map is plotted in four rotational phases.
              }
         \label{fig_diall12}
\end{figure*}

\subsubsection{Doppler images from the 2005 CFHT spectra}
In Fig.\,\ref{fig_di12}, the two subsequent Doppler maps for the 2005 season (S1, S2) are plotted in orthographic projection in the quadrant phases of the rotation. The corresponding line profile fits are plotted in Figs.\,\ref{fig:proffits_s1}-\ref{fig:proffits_s2}. Both temperature maps show mostly cool spots at similar locations, accounting for the reliability of the image reconstruction. Indeed, most of the cool surface features appear on the combined Doppler map derived from all of the spectra (S1+S2); see Fig.\,\ref{fig_diall12}.
%In accordance with \cite{2006MNRAS.367.1699H},  

The strongest feature, a cool, high latitude spot on the hemisphere of the visible pole of $\Delta T>-$1000\,K temperature contrast relative to $T_{\rm eff}$ is shown on both S1 and S2 maps at around 0.45 phase, centered at $\beta\approx50$\degr. In addition, this large spot is getting more compact from S1 to S2. A mid-latitude spot is seen on the opposite (lower) hemisphere at $\approx$0.3 phase. Its location, however, is a bit shifted from S1 to S2 towards the invisible pole, or alternatively, its extension decreases. %probably as an indication of surface differential rotation, which will be discussed further in Sect.\,\ref{subsect:diffrot}.
A weak cool spot is seen on S1 at $\approx$0.13 phase, elongated from 30\degr\ latitude down to the equator and even below to $-$30\degr. Such an elongated shape, however, may be the result of mirroring due to the high inclination. In deed, its antitype on S2 is less contrasted but more circular shaped and confined to the upper hemisphere, i.e., less mirrored. Another less contrasted cool spot is seen on S2 centered at 0.83 phase and 20\degr\ latitude, which, again, has a weaker but more extended precursor on S1.

\begin{figure*}%[]
            
            \vspace{2.1cm}
            \Large{S3}
            \vspace{-2.7cm}
            
            \hspace{0.5cm}\includegraphics[width=0.98\textwidth]{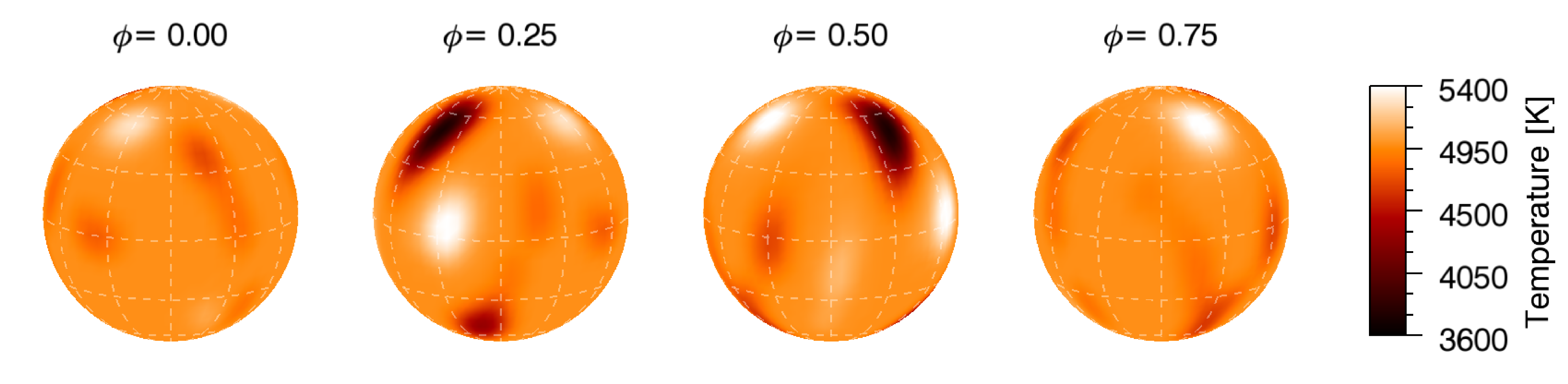}
    
            \vspace{2.1cm}
            \Large{S4}
            \vspace{-2.7cm}
            
            \hspace{0.5cm}\includegraphics[width=0.98\textwidth]{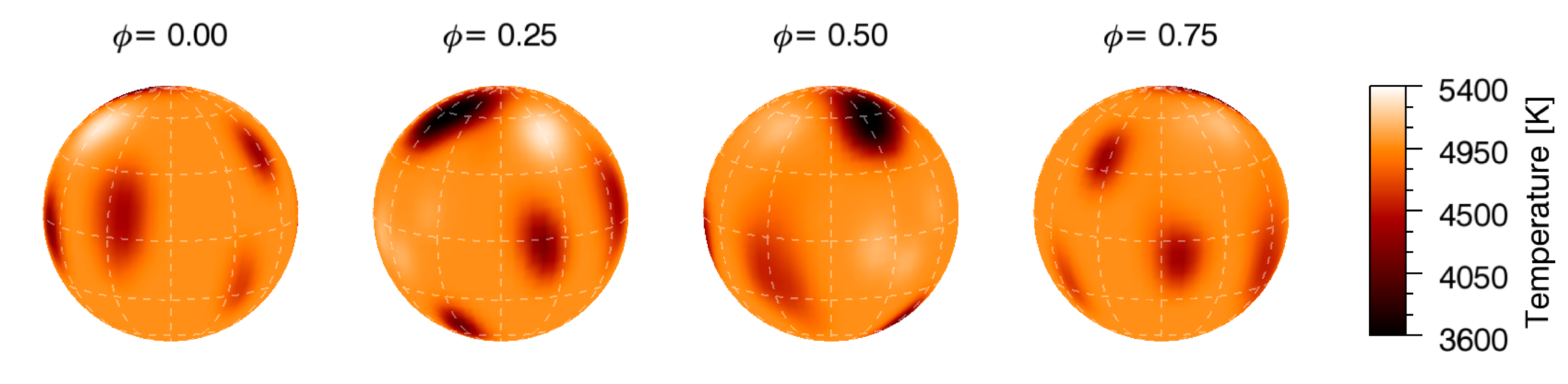}
            
      \caption{Two Doppler images of V471\,Tau for the second observing run at the turn 2014/2015 (S3 and S4). The surface temperature reconstructions are plotted in four rotational phases.
              }
         \label{fig_di34}
\end{figure*}
\begin{figure*}%[]
%    \resizebox{\hsize}{!}
             
            \vspace{2.1cm}
            \Large{S3}
            
            \hspace{0.1cm}\Large{+} 
            
            \Large{S4}            
            
            \vspace{-3.2cm}
            \hspace{0.5cm}\includegraphics[width=0.98\textwidth]{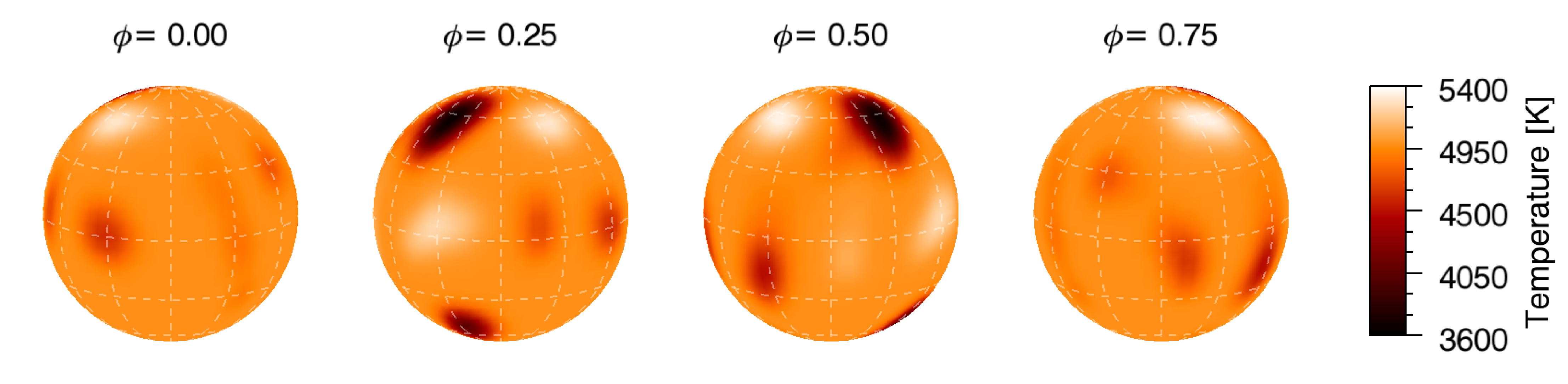}
      \caption{Combined Doppler image of V471 Tau derived from all the spectra in the second observing run at the turn 2014/2015 (S3+S4). The surface temperature map is plotted in four rotational phases.
              }
         \label{fig_diall34}
\end{figure*}

Some weak bright features are also seen, but their contrast hardly reaches $\Delta T\approx+$200\,K and their location and shape are more changeful when comparing S1 image against S2. Their presence is even less salient on the average (S1+S2) image except the one at 0.3 phase. This feature, however, raises reliability issues. We know that spurious bright features can appear as an artifact of Doppler imaging coming from insufficient phase coverage \citep[e.g.][]{lindborg1}. The phase coverages for both S1 and S2 were fairly good, the only remarkable phase gaps were between 0.288-0.500 and 0.171-0.318 in S1 and S2 datasets, respectively. This may explain the bright spot at 0.3 phase on the S2 reconstruction and account for some minor inconsistencies between S1 and S2 around these phase gaps. All in all, we think that those bright spots are mostly artifacts.

\subsubsection{Doppler images from the 2014/15 CFHT spectra}
Doppler images for the 2014/15 observing season (S3, S4) are presented in Fig.\,\ref{fig_di34}; for the corresponding line profile fits see Figs.\,\ref{fig:proffits_s3}-\ref{fig:proffits_s4}. Again, the most dominant cool features in two subsequent maps appear at similar locations. However, neither the weaker cool features nor the bright features are particularly consistent. For example, bright features are more apparent in the S3 reconstruction, compared with S4. This is not surprising when keeping in mind that due to the unfavourable data distribution the selected spectra cover $\approx$20 days for S3 and ten days for S4. Moreover, the time gap between the mean HJDs of S3 and S4 is $\approx$15 days, therefore incoherences between the two images are more visible. The combined Doppler map derived from all of the spectra (S3+S4) in Fig.\,\ref{fig_diall34} indicates that the most coherent and most striking feature is the high latitude ($\beta\approx55$\degr) cool spot at around $\approx$0.4 phase, recalling the S1 and S2 reconstructions. This cool spot may be coupled to the orbital phase, as suggested in \citet[][]{1991A&A...251..183S}. We conclude that the weak and inconsistent features in the S3 and S4 reconstruction, being cool or bright, can either be the result of imperfections of the reconstruction or come from the unfavourable conditions of the data distribution.

\subsection{Differential rotation}
\label{subsect:diffrot}

Surface differential rotation can be measured by longitudinally cross-correlating consecutive Doppler images \citep{donati_ccf,2004A&A...417.1047K}, and fitting the latitudinal correlation peaks in the resulting cross-correlation function (ccf) map by a quadratic rotational law:  
\begin{equation}\label{drlaw}
    \Omega(\beta)=\Omega_{\mathrm{eq}}-\Delta\Omega\sin^2\beta,
\end{equation}
where $\Omega(\beta)$ is the angular velocity at $\beta$ latitude, $\Omega_{\mathrm{eq}}$ is the angular velocity of the equator, and $\Delta\Omega=\Omega_{\mathrm{eq}}-\Omega_{\mathrm{pole}}$ gives the difference between the equatorial and polar angular velocities. With these, the dimensionless surface shear parameter $\alpha_{\mathrm{DR}}$ is defined as $\alpha_{\mathrm{DR}}=\Delta\Omega/\Omega_{\mathrm{eq}}$, leading the following form:
\begin{equation}\label{eq:drlaw}
    \Omega(\beta)=\Omega_{\mathrm{eq}}(1-\alpha_{\mathrm{DR}}\sin^2\beta).
\end{equation}

\begin{figure}%[]
    \includegraphics[width=\columnwidth]{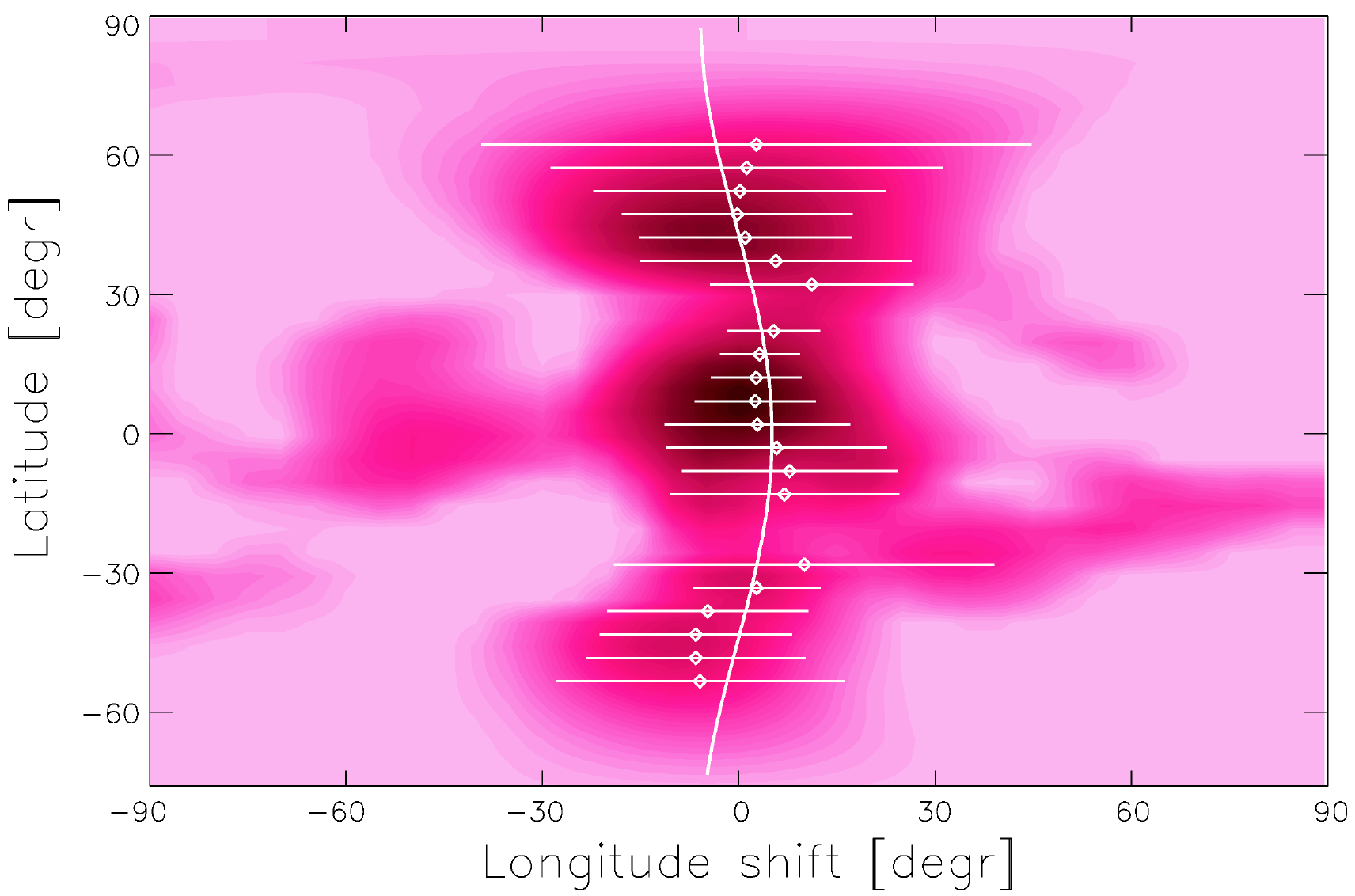}
      \caption{Average cross-correlation function map derived from the subsequent Doppler images plotted in Fig.\,\ref{fig_di12} and Fig.\,\ref{fig_di34}. Black represents strong correlation while light pink indicates no correlation. The quadratic sine fit (solid line) applied to the ridge of the correlation pattern (dots with error bars) suggests a solar-type rotational law with a very weak shear of  $\alpha_{\mathrm{DR}}=0.0026\pm0.0006$. See Sect.\,\ref{subsect:diffrot} for details.
      }
        \label{fig_ccf}
\end{figure}

\begin{figure}[th]
    \includegraphics[width=\columnwidth]{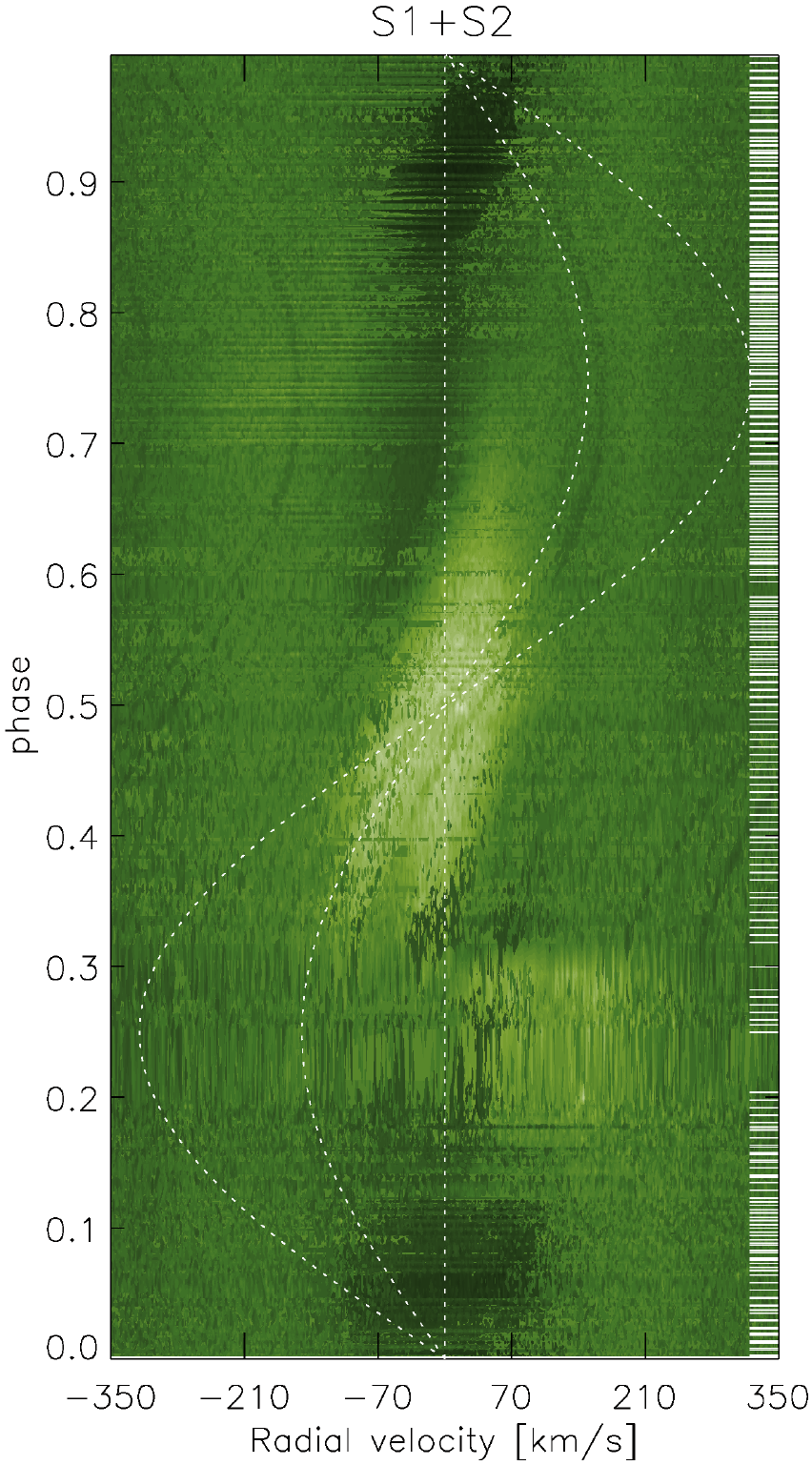}
      \caption{Dynamic H$\alpha$ spectrum from the 2005 data (S1+S2). The individual 1-D H$\alpha$ spectra along the orbital phase are plotted in the rest frame of the K star. Dark green corresponds to deep absorption, while light yellow indicates strong emission. The large amplitude ($\approx$320\,km\,s$^{-1}$) sine curve marks the radial velocity of the white dwarf, the smaller amplitude ($\approx$150\,km\,s$^{-1}$) sinusoid is the radial velocity of the centre of mass of the binary system. Tick marks on the right indicate the phases of the observations.}
        \label{fig:halpha12}
\end{figure}
\begin{figure}[th]
    \includegraphics[width=\columnwidth]{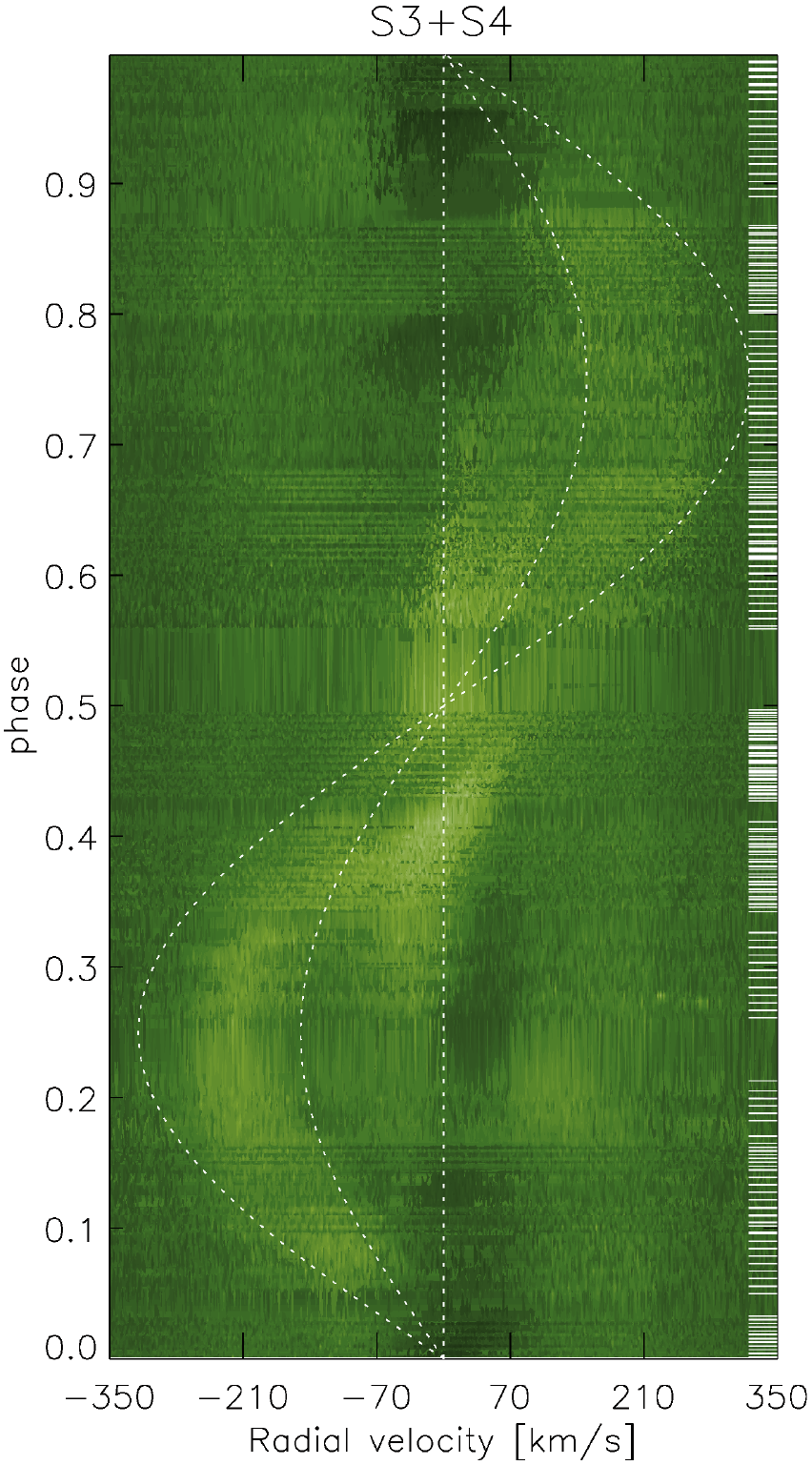}
      \caption{Dynamic H$\alpha$ spectrum from the 2014/2015 data (S3+S4). Otherwise as in Fig.\,\ref{fig:halpha12}.}
        \label{fig:halpha34}
\end{figure}

In our case two ccf maps can be composed, one for the two consecutive maps in the first observing season (S1, S2) and another one for the two maps in the second season (S3, S4). After the necessary normalization procedure the two ccf maps are combined to get an average ccf map \citep[for the applied method dubbed \texttt{ACCORD} see e.g.,][]{kovari_zetand,2015A&A...573A..98K}.
The correlation pattern in the ultimate average ccf map (Fig.\,\ref{fig_ccf}) is fitted by a rotational law according to Eq.\,\ref{eq:drlaw}. The resulting fit indicates a very weak solar-type differential rotation, with $\Omega_{\mathrm{eq}}=691.57\pm0.13$\degr/day and $\alpha_{\mathrm{DR}}=0.0026\pm0.0006$ shear parameter, i.e., almost solid body rotation, in a fair agreement with the rough estimation from the \emph{K2} photometry (see Sect.\,\ref{sect_k2diffrot}). From the fitted $\Omega_{\mathrm{eq}}$ and $\alpha_{\mathrm{DR}}$ values we get $\beta_{\mathrm{cor}}\approx43$\degr\ for the co-rotating latitude, i.e., where rotation is synchronized exactly to the orbit. 

%-----------------------------------------------------------------

\section{Chromospheric and coronal activity}

\subsection{Variability of the H$\alpha$ line profile}\label{sect:halpha}

H$\alpha$ line profile variation is capable to unfold chromospheric activity and presumably associated mass motions like CMEs. From the available time series of spectra we create time-slice plots for both observing runs. The resulting dynamic spectra are shown in Figs.\,\ref{fig:halpha12}-\ref{fig:halpha34}. The plot for the first observing run (S1+S2) indicates absorption around 0.0 phase when the white dwarf is near or in eclipse. On the other hand, a strong emission develops between around 0.35-0.6. Clearly, the peaks of the absorption and the emission profiles stay within the $\pm$91\,km\,s$^{-1}$ projected equatorial velocity range of the K star, suggesting that the sources are related to some surface features. Especially, the emission around the secondary eclipse may indicate relation with the dominant cool photospheric spot at around 0.45 phase (cf. Fig.\,\ref{fig_di12}). There may also be some extra emission on or over the receding side of the K star between 0.2-0.3 phase, however, this part of the data has the lowest quality and even the phase coverage is not satisfactory, therefore we cannot make a definite statement about this feature.

The H$\alpha$ behaviour from the 2014/2015 data (S3+S4) in Fig.\,\ref{fig:halpha34}) shows similarities to the dynamic H$\alpha$ spectrum from the 2005 data, but quite different features are also present. Similar features are the absorption centered around 0.0 phase and the emission between 0.35-0.65 phase, this latter less conspicuous but still around the phase of the most prominent cool spot (see Fig.\,\ref{fig_di34}). Being within the radial velocity range of the K dwarf, it is very likely that these features are localized on the surface of the red dwarf. However, in addition to the similarities there is a very different emitting phenomenon located between the centre of mass of the system and the white dwarf. This emission is stronger in the first half of the orbital phase, but still visible during the second half. This suggests the presence of a clump of emitting plasma corotating with the system in the vicinity of the inner Lagrangian (L1) point, as already reported by \citet{1991ApJ...378L..25Y}.
We note finally, that there may be some extra emission on the receding side of the K star around $\phi$=0.2, recalling Fig.\,\ref{fig:halpha12}, but the phase coverage is not satisfactory around and the emission peaks are not so strong, therefore, again, we could only guess about its reliability, nature and possible origin.

For further discussion about the peculiar emission from beyond the L1 point see Sect.\,\ref{L1emission}.

\subsection{X-ray emission from the K dwarf}\label{sect:xray}

The \emph{ROSAT} pointed observations from 1991 were already analyzed by \citet{1998MNRAS.297.1145W} who focused on the second, longer exposure that we also use here. The full wavelength coverage of the PSPC detector is 0.1-2.5\,keV, i.e. the softer part is dominated by the emission from the white dwarf. \citet{1998MNRAS.297.1145W} presented the light curve of the observation in a soft (0.1-0.4\,keV) and a hard (0.4-1.2\,keV) band. The soft band clearly shows the (partly covered) eclipse, which is not seen in the hard band, confirming that the harder X-ray emission stems from the K dwarf only. Both bands show variability, but no large flares. The \emph{ROSAT} data are, however, affected by many data gaps due to the spacecraft orbit. Therefore, we simply use the average flux from this observation as the quasi-quiescent state for this epoch. The 0.1-2.5\,keV flux from the K dwarf only (determined by spectral modelling in \citet{1998MNRAS.297.1145W}) amounts to $(4.0\pm0.2)\times10^{-12}$\,erg\,cm$^{-2}$\,s$^{-1}$.

In 1996, V471\,Tau was observed by \emph{ASCA}. This observation is described and analyzed in detail by \citet{2003ApJ...597.1059S}. Their light curve from the combined SIS and GIS data in the energy range 0.3-10\,keV shows some variability which may be due to small flare events, but the number of potentially flaring data points is small compared to the full light curve. \citet{2003ApJ...597.1059S} used the whole observation for spectral modeling (due to the rather low count rate) and obtained an average flux of $2.4\times10^{-12}$\,erg\,cm$^{-2}$\,s$^{-1}$ valid for the 0.5-10\,keV range, which we adopt here as well.

One year later, V471\,Tau was observed with the \emph{ROSAT} High Resolution Imager (HRI). This instrument has an energy range of 0.1-2.5\,keV, but due to its negligible energy resolution the data cannot be binned in energy to remove the contribution from the white dwarf. Therefore, we extracted the light curve of the whole observation using the \textit{xselect} task from the \textit{HEASoft}\footnote{\url{https://heasarc.gsfc.nasa.gov/docs/software/heasoft/}} package. This observation covered a part of one eclipse. Therefore, we used \textit{xselect} to extract an image for the time interval of the eclipse only (thus containing only emission from the K dwarf) and determined the count rate with the \textit{ximage/sosta} task, which performs also background subtraction and corrects for vignetting, deadtime, and PSF. Although the full light curve did not show any large flares, the short duration of the usable time interval makes this data point more uncertain. 

The \emph{Chandra} observations from 2002 were already analyzed in several studies \citep{2003ApJ...594L..55D,2004A&A...427..667N,2005ApJ...621.1009G}. The HRC-S/LETG instrument provides high-resolution spectroscopy in the 0.07-7.29\,keV range. The light curve shown by \citet{2005ApJ...621.1009G} does not include any strong flares, so the average flux of the observation was adopted. This was taken from \citet{2004A&A...427..667N} who extracted the flux in the smaller 5.15-38.19\,\AA\ wavelength band (quoted luminosity of $1.1234\times10^{30}$\,erg\,s$^{-1}$), i.e. with negligible contribution from the white dwarf.

The first \emph{XMM Newton} observation from 2004 was analyzed using the \emph{XMM Newton} Scientific Analysis System (SAS) Version 17.0.0. We only use data from the EPIC/PN detector, as the simultaneous EPIC/MOS and RGS data have lower S/N ratio. The SAS task \textit{epproc} was used to create an event list from the raw odf file for the EPIC/PN detector. We extracted source and background regions in \textit{ds9} and use the SAS tasks \textit{evselect} to create the light curves and \textit{epiclccorr} the perform background subtraction and all necessary corrections (vignetting, deadtime, etc.). We use the energy band 0.3-10\,keV to minimize potential contamination from the white dwarf at lower energies, and because we use this energy range for the common analysis of all instruments. This observation is affected by a large flare event previously analyzed by \citet{2008MNRAS.387.1627P} and significant variability. Therefore, we needed to use a restricted time range that most likely resembles a quasi-quiescent state. We extract the last 18783\,s of the observation which seems to be free of any trends, although there is still strong scatter in the light curve.

In 2015, V471\,Tau was observed by \emph{Swift}. We inspected the light curves from the XRT instrument created by the online tool of the \emph{Swift} Science Center\footnote{\url{https://www.swift.ac.uk/user_objects/}} \citep{2007A&A...469..379E,2009MNRAS.397.1177E}. There are no strong flares in the data, so we simply adopt the averaged count rate from all observations in the 0.3-10\,keV band.

The second \emph{XMM Newton} observation from 2019 was processed an analyzed in the same way as described before. This observation was also affected by a large flare and strong variability, which required to extract the first 9381\,s only, which seems to display a rather constant count rate without trends.

\subsection{Evolution of the X-ray luminosity}
To compare the X-ray emission from many different instruments requires to either convert the fluxes taken from the literature to a common energy range or to obtain the appropriate count-to-flux conversion factors for the different instruments. We use the tool \textit{WebPIMMS}\footnote{\url{https://heasarc.gsfc.nasa.gov/cgi-bin/Tools/w3pimms/w3pimms.pl}} to perform the necessary conversions. This requires input of the Galactic hydrogen absorption and a spectral model for the source.

\begin{table*}%[]
\caption{Spectral fits to the quiescent parts of the two \emph{XMM Newton} observations from 2004 and 2019 and the \emph{Swift} data from 2015. Errors refer to the 90\% confidence intervals returned by \textit{Xspec}. The last column gives the goodness-of-fit and the degrees of freedom.}
\label{xspec}
\begin{tabular}{crrrrrrrr}
\hline\noalign{\smallskip}
Data & $kT_1$ & $EM_1$ & $kT_2$ & $EM_2$ & $Z$ & $T_\mathrm{av}$ & $EM_\mathrm{tot}$ & $\chi^2_\mathrm{red}$ (DOF) \\
 & [eV] & [$10^{52}$ cm$^{-3}$] & [eV] & [$10^{52}$ cm$^{-3}$] & [$Z_\sun$] & [MK] & [$10^{52}$ cm$^{-3}$] &  \\
\hline\hline\noalign{\smallskip}
\smallskip
XMM 2004 & $0.34^{+0.05}_{-0.02}$ & $3.1^{+0.2}_{-0.2}$ & $0.96^{+0.02}_{-0.02}$ & $6.7^{+0.4}_{-0.4}$ & $0.18^{+0.02}_{-0.02}$ & 8.86 & 9.9 & 1.503 (58) \\
\smallskip
XMM 2019 & $0.36^{+0.04}_{-0.03}$ & $2.6^{+0.3}_{-0.3}$ & $1.01^{+0.02}_{-0.02}$ & $6.0^{+0.4}_{-0.4}$ & $0.18^{+0.02}_{-0.02}$ & 9.48 & 8.6 & 1.043 (58) \\
\smallskip
Swift 2015 & $1.03^{+0.15}_{-0.08}$ & $11.9^{+2.5}_{-2.3}$ & -- & -- & $0.13^{+0.07}_{-0.05}$ & -- & -- & 0.999\tablefootmark{a} (132) \\
\hline
\end{tabular}
\tablefoot{
\tablefoottext{a}{Pearson $\chi^2_\mathrm{red}$ due to use of Cash statistics for fitting.}
}
\end{table*}

To constrain the plasma parameters of the K dwarf in V471\,Tau we perform spectral modeling of the two \emph{XMM Newton} data sets and \emph{Swift}. For \emph{XMM Newton}, we extract source and background spectra of the selected quasi-quiescent time ranges using \mbox{\textit{evselect}} and create the required \textit{rmf} and \textit{arf} files using \mbox{\textit{rmfgen}} and \mbox{\textit{arfgen}}. We use \mbox{\textit{specgroup}} to group the files and select a minimum of 20 counts per bin and not oversampling the intrinsic energy resolution by more than a factor of three. We fit the spectra using \textit{Xspec}\footnote{\url{https://heasarc.gsfc.nasa.gov/xanadu/xspec/}} with a model including Galactic absorption and a sum of APEC \citep{2001ApJ...556L..91S} plasma models. As the Galactic absorption is not very well constrained by a fit to energy ranges $>0.3$\,keV we fix the H column density towards V471\,Tau to its measured value of $N_\mathrm{H}=1.62\times10^{18}$\,cm$^{-2}$ \citep{2005ApJS..159..118W}. This has a negligible effect on the results, as this value is low because the star is nearby. For the fitting process, we restrict the energy range further to 0.3-3.9\,keV due to low signal at higher energies. At both epochs, we find best fits with two APEC plasma components with the same coronal abundances (Table\,\ref{xspec}). The results are broadly consistent with each other, as well as other spectral analyses in the literature.  We also tried to fit the \emph{Swift} spectrum obtained from the online tool \citep{2009MNRAS.397.1177E}, but due to the small number of counts it was not possible to fit a 2-T plasma model. An acceptable fit was obtained for a 1-T model using Cash statistics \citep{1979ApJ...228..939C}, yielding a temperature comparable to the hotter component of the more reliable \emph{XMM Newton} fits and an emission measure similar to the total one obtained for \emph{XMM Newton}. All fits predict subsolar coronal abundances ($Z\sim0.2Z_\sun$), consistent with previous studies. Table\,\ref{xspec} summarizes the plasma parameters obtained from the \textit{Xspec} fits, together with the calculated average emission measure-weighted temperature $T_\mathrm{av}$ and total emission measure $EM_\mathrm{tot}$ for the 2-T fits.

\begin{figure}%[]
    \includegraphics[width=\columnwidth]{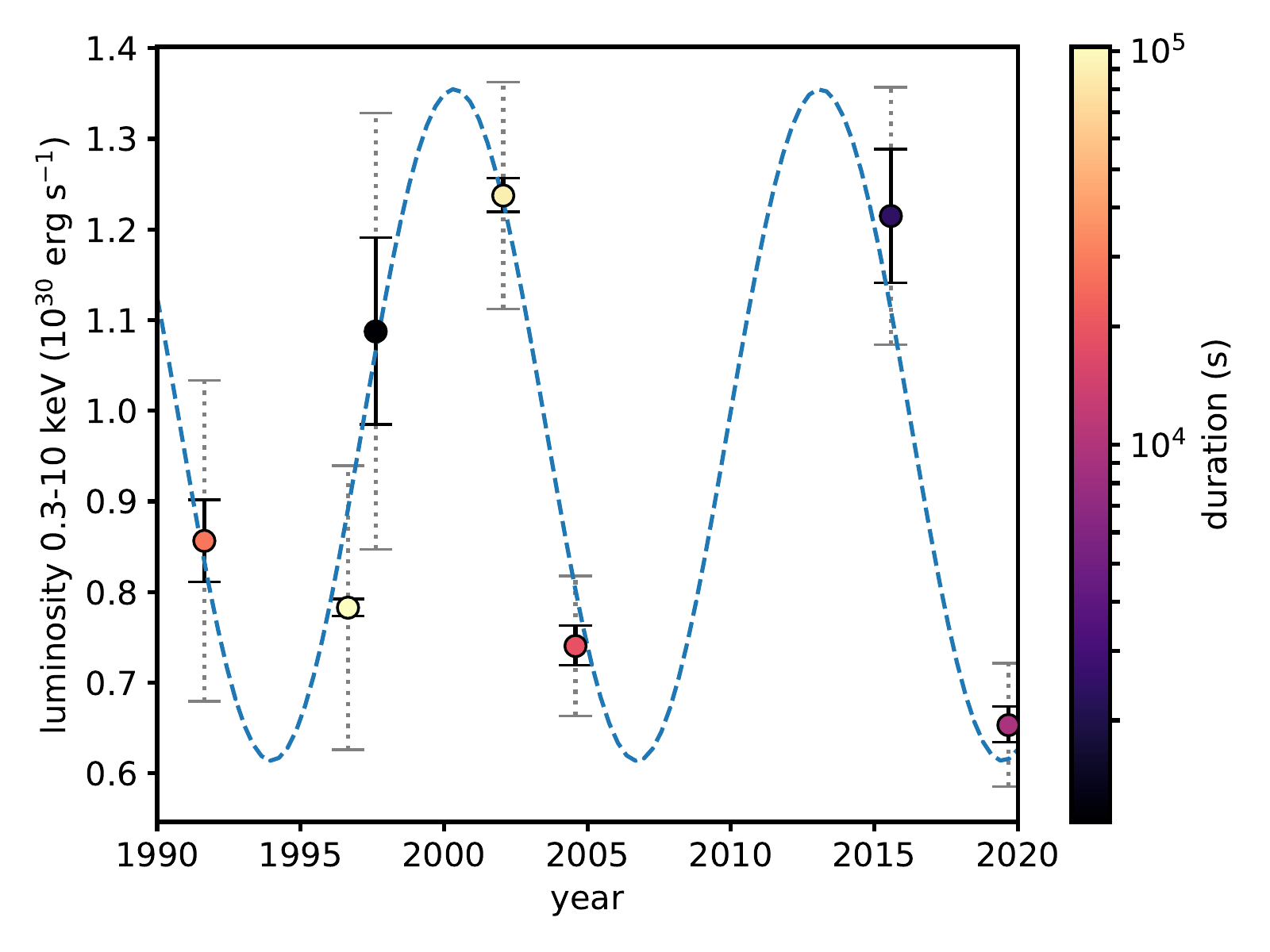}
      \caption{Evolution of the X-ray luminosity of the K dwarf in V471\,Tau from the observations summarized in Table\,\ref{xrayobs}. The duration of the selected quiescent ranges is color coded. Solid error bars denote flux errors (see text), dotted bars additionally the estimated cross-calibration uncertainty. The blue dashed line indicates the best fit with a sinusoidal function with a period of $\sim$12.7\,yr.}
        \label{fig_lx}
\end{figure}

As we do not find significant differences of coronal plasma parameters from the analyzed observations, we adopt a common parameter set for all flux conversions. For simplicity, we use a 1-T APEC model in \textit{WebPIMMS} with a mean temperature of 8.9\,MK ($\log T=6.95$). We fix the Galactic $N_\mathrm{H}$ absorption to the value from \citet{2005ApJS..159..118W} and the coronal metal abundance to $0.2Z_\sun$. Using a 2-T model with parameters similar to those obtained in the \emph{XMM-Newton} fits does not change the results significantly. To estimate uncertainties for the conversion factors, we run \textit{WebPIMMS} with $\log T=6.9$ and $\log T=7.0$, corresponding to an uncertainty of the mean temperature by about 10\%, which is slightly larger than what we calculate from the \emph{XMM-Newton} spectra (3-4\%). This results in uncertainties on the conversion factors by $\lesssim$3\%.

After converting the fluxes taken from the literature and the count rates of the specific instruments to a common energy range of 0.3-10\,keV, we use the distance from Table\,\ref{table_pars} to convert the fluxes to luminosities. The results are shown in Fig.\,\ref{fig_lx}. The displayed uncertainties on the X-ray luminosities include the statistical uncertainties of the flux/count rate measurements, the flux conversion factors, and the distance. The X-ray luminosity of the K dwarf shows variations over the years, with the maximum data point being 90\% higher than the minimum one. As there are only seven data sets spanning almost 30 years, some with exposure times for the quasi-quiescent time intervals as short as $\sim$20\,min, a detailed investigation of a possible cycle is not feasible. Moreover, the data stem from a variety of different instruments and cross-calibration issues are expected to be present \citep[see the detailed discussion in][]{2017MNRAS.464.3281W}. To account for such instrument cross-calibration uncertainties, we add 10\% for \emph{Chandra}, \emph{XMM-Newton} and \emph{Swift} data, as well as 20\% for the older \emph{ROSAT} and \emph{ASCA} data to the luminosity uncertainties \citep{2002astro.ph..3311S,2011A&A...525A..25T}.

Despite the limitations of the available data, we fitted a sinusoidal function to the X-ray luminosities in Fig.\,\ref{fig_lx}. The blue dashed line in the plot corresponds to a period of about 12.7\,yr, which can be interpreted as activity cycle. We stress again, however, that the uneven spacing of the observations, the small number of data points, as well as uncertainties of instrumental cross-calibration, do not allow a definitive conclusion about the existence of an X-ray cycle and its properties.

\section{Discussions}
\label{sect:disc}

\subsection{Weak differential rotation and active longitude}

From our Doppler reconstructions we have measured a very weak solar-type surface differential rotation for the K star (Sect.\,\ref{subsect:diffrot}), 
Such a weak shear, however, is not surprising, as we essentially expect something similar for fast rotating late-type dwarf stars \citep[see the empirical study by][and their references]{2017AN....338..903K}. In addition, synchronization due to tidal dissipation in close binaries may suppress the rotational shear \citep{1982ApJ...253..298S}. In \citet{2006MNRAS.367.1699H} authors reported almost rigid body rotation for V471\,Tau with $\alpha_{\rm DR}$ of $0.0001\pm0.0005$ shear coefficient. In comparison, our result of $\alpha_{\rm DR}=0.0026\pm0.0006$ is still weak, but an order of a magnitude stronger shear based on using much more data and our well-established cross-correlation technique \texttt{ACCORD} \citep[for some applications and tests see, e.g.][and their references]{2004A&A...417.1047K,kovari_zetand,2014IAUS..302..198K,2015A&A...573A..98K}. This technique amplifies independent cross-correlation signals from different epochs to multiply the credibility and reliability of the result. We note finally, that our larger shear coefficient is also supported by the short term photometric period changes (see Sect.\,\ref{sect_k2diffrot}).

The Doppler reconstructions performed in our study indicate that the most dominant surface spot located around 0.40-0.45 phase at $\beta\approx50$-55\degr\ latitude may be a permanent feature (see Figs.\,\ref{fig_di12}-\ref{fig_di34}). In the Doppler imaging study by \citet{1995AJ....110.1364R} authors have found that a dominant cool spot appeared at the longitude facing the white dwarf in all their four Doppler maps (1992 September, October, December and 1993 December). Moreover, the Doppler image from 2002 November \citep{2006MNRAS.367.1699H} has also indicated that the most prominent spot was directed toward the white dwarf.

Asymmetric tidal distortion may significantly alter internal dynamos of rapidly rotating convective stars in close binaries  \citep{2002A&A...392..535R}, which can result in an active longitude around the sub-white dwarf point in the K dwarf component. Such a non-axisymmetric nature of spot activity is not unusual in close binary systems \citep{2006Ap&SS.304..145O}. However, it has also been shown that the condition for the excitation of stable, non-axisymmetric fields is weak differential rotation \citep{1982ApJ...253..298S,1994A&A...281...46R}, which we firmly confirmed in the case of V471\,Tau.

\subsection{Effective temperature rise around $\phi$=0.5 phase}\label{Teffrise}

The $\approx$100\,K variation of the surface temperature of the K dwarf along the orbital phase (see Fig.\,\ref{fig:teffvar}) may be interpreted as the irradiation effect of the white dwarf. In fact, we have no means to calculate what fraction of the radiative energy from the white dwarf is converted to thermal reprocessing in the photosphere of the K star. However, according to \citet[][their Table 2]{1988ApJ...334..397Y} the estimated ratio of the total luminous energy intercepted from the white dwarf to the bolometric output of the K star is $\approx$0.6\%, i.e., below the detection limit of 3\% assumed for such a heating mechanism to be efficient in the photosphere. A similar conclusion has been drawn by \citet{2001ApJ...563..971O}, who estimated that the luminosity of the K star would increase only by 1\% due to the irradiation from the white dwarf, which would explain only $\approx$12\,K temperature rise of the sub-white dwarf hemisphere. In contrast, the effective temperature change of  $\approx$100\,K in Fig.\,\ref{fig:teffvar} would assume $\approx$8\% luminosity difference. At this point we speculate that the photospheric temperature rise around $\phi$=0.5 may rather be associated with increased magnetic activity at the sub-white dwarf hemisphere of the K star. It has been learned that faculae-dominated active stars tend to become brighter when their magnetic activity level increases \citep{1990Sci...247...39R}. We believe that the hemispheric difference in magnetic activity observed in V471\,Tau can have a significant effect on the overall hemispheric temperature difference as well \citep[cf. e.g.][]{2008A&A...478..883G,2016A&A...589A..46S}.
That is, small-scale photospheric bright faculae, permanently present around the active longitude facing the white dwarf, may infer such a rise in the effective temperature.
%The permanent active longitude on the K star  would imply that small scale photospheric bright features (faculae) have an overall significant impact on the hemispheric effective temperature. 

\subsection{Flare statistics}\label{sect:disc:flares}
\begin{figure}%[]
    \includegraphics[width=\columnwidth]{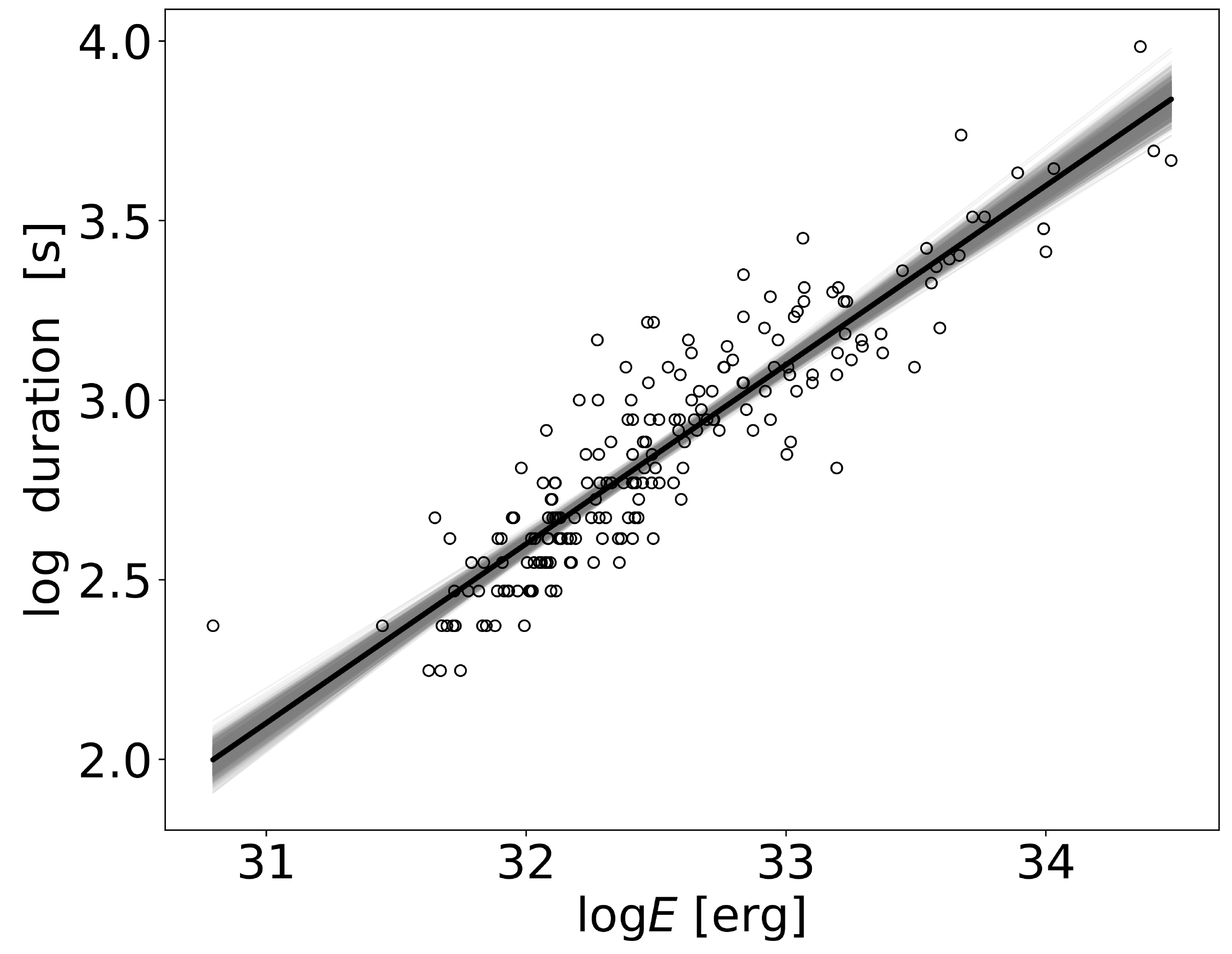}
      \caption{Flare duration vs. flare energy from the \emph{K2} data of V471\,Tau.   The best fit indicates a power-law  relationship with 0.5$\pm$0.02 index.}
        \label{fig:flaredur}
\end{figure}

We find that the 198 flares detected in the 71-day long \emph{K2} data of V471\,Tau are distributed randomly over the rotational phase, i.e. no significant phase dependency was found in the flare occurrence.
The derived flare energies in Sect.\,\ref{sect:flares} range over four magnitudes. Recently, \citet{2020A&A...641A..83K} have concluded that, in principle, differences between flare energies are due to size effect, regardless of energy range or spectral type \citep[cf.][]{2015MNRAS.447.2714B}. Accordingly, flare duration is supposed to increase with flare energy. In Fig.\,\ref{fig:flaredur} we plot $\log\Delta t$ flare duration as a function of $\log E_{\rm f}$ flare energy. The relationship on the log-log plot fitted by a linear function yields a 0.50$\pm$0.02 slope. In comparison, \citet{2015EP&S...67...59M} suggested a generalized function between flare duration and flare energy with a power-law index (slope) of $\approx$1/3.

It has been suggested \citep[see e.g.][]{1988ApJ...330..474P,1991SoPh..133..357H,1998ASPC..154.1560S} that the more numerous small flaring events have an important role in heating the plasma in the transition region and the corona. This is supported by the power-law index of $\alpha\gtrsim2.0$ derived from flare statistics of magnetically active M dwarf stars \citep[e.g.,][etc.]{2019A&A...622A.133I}. In comparison, $\alpha=1.7$ was obtained for 27 young K 5-7 stars in the Orion Nebula Cluster \citep{2005ApJS..160..423W}. However, no such a difference was found from the flare statistics for 548 M and 343 K dwarf stars ($\alpha=1.82$ and 1.86, respectively) in \citet{2019ApJ...873...97L}. Nevertheless, the $\alpha$ index of 1.8 obtained for V471\,Tau (see Fig.\,\ref{fig_ffd}) suggests that the dissipating energy of the flares should have a significant contribution in heating the upper atmosphere.

%If α≥2,  the  total energy  released  in  flares  can  be  arbitrarily  high  (Güdel  et  al.2002)
%+ X luminosity
%small flares: main source for heating the corona

%As for the reliability from another aspect, it is known that at $i=90$\degr\ inclination degeneracy occurs, i.e. the imaging process cannot make a distinction between the two hemispheres, therefore they are mirrored. But this kind of mirroring effect can develop even at relatively high inclinations like $i=76$\degr in our case, making the image reconstruction less reliable.

\subsection{Inter-binary H$\alpha$ emission during activity maximum}\label{L1emission}

%+ Active regions represent the brightest coronal areas. The reason behind this is that in active regions along closed magnetic loops hot gas is trapped, which radiate in X-rays.
%+ X-ray flux is well correlated with the magnetic flux density (Schrijver et al., 1992; Schrijver and Title, 2005)

The cycle period of the sinusoid fitted to the X-ray luminosity variation in Fig.\,\ref{fig_lx} is comparable to the 13\,yr period identified in timing residuals by \citet{2015ApJ...810..157V} and a possible $\approx$10\,yr periodicity from an activity cycle mentioned by \citet{2007AJ....134.1206K}. Moreover, \citet{2007AJ....134.1206K} stated that during observations with \emph{MOST} in Dec 2005 the star was much less spotted than in the previous observations of \citet{2006MNRAS.367.1699H} (from Nov 2002) and had only little photometric variability, which the authors interpreted as the evidence that the K dwarf was close to an activity minimum. This is consistent with our fit to the X-ray data.

According to Fig.\,\ref{fig_lx}, about four months ahead of the epochs of the S1-S2 Doppler reconstructions, in 2004 the X-ray luminosity of V471\,Tau was around its minimum. While about six months after the epochs of the S3-S4 Doppler images, in 2015 the X-ray luminosity was much higher, according to the data point from \emph{Swift}. Oddly, such a large difference is not seen when comparing the overall spottedness in the S1-S2 images with that of S3-S4, which are quite similar, despite the $\approx$9-year time gap between them.  However, the H$\alpha$ line profile variability (see Figs.\,\ref{fig:halpha12}-\ref{fig:halpha34}) showed indeed different behaviour in the two observing runs. The extra H$\alpha$ emission from the inter-binary space (see Fig.\,\ref{fig:halpha34}) might have something to do with the enhanced X-ray luminosity in 2015, i.e., around activity maximum. In deed, this is supported by the result in \citet[][see their Fig. 3]{1991ApJ...378L..25Y} who also found external H$\alpha$ emission component, originated from somewhere in the inter-binary space beyond the centre of mass of the system, expanding towards the white dwarf, definitely reminiscent of our result plotted in Fig.\,\ref{fig:halpha34}. Moreover, their observations were obtained in Nov 3-4, 1988 and Jan 29-30, 1989, i.e., close to the suspected maximum of the activity cycle ended in 1994 (according to our Fig.\,\ref{fig_lx}). As a further support see also \citet[][]{2002A&A...392..535R} for the secular H$\alpha$ variability in 1985, 1990 and 1992.

We expect that during the magnetic cycle, around activity maximum more and/or more extended magnetic loops develop high above the surface, reaching one stellar radius \citep{1986ESASP.263..197G}, and this way more cool material enters the upper atmosphere in trapped clumps, similar to solar prominences. Large prominence-like condensations of cool material were reported from H$\alpha$ observations of AB\,Dor \citet{1989MNRAS.238..657C}, a single K0 dwarf, rotating as rapidly as V471\,Tau.   Normally, such a clump of cool material appears in absorption in H$\alpha$, exactly what we actually see on the far side of the K star around zero phase. However, it can get into emission when heated due to the UV radiation from the white dwarf. The physical process behind heating proposed first by \citet{1988ApJ...334..397Y} involves fluorescence-induced H$\alpha$ emission \citep[see also][]{1991AJ....102.2079B}. This scenario could explain the H$\alpha$ line variability both around the activity minimum (Fig.\,\ref{fig:halpha12}) when there are no extended magnetic loops, and around maximum (Fig.\,\ref{fig:halpha34}) when extended loops with cool material trapped in them are present.
Still, we caution that our ground-based spectroscopic data and the space observations did not overlap exactly and the available space data is limited. Therefore our conclusion about such a link between the peculiar H$\alpha$ emission from the vicinity of the L1 point and the proximity of the activity cycle maximum may raise some criticism.

%Moreover, we have only seven X-ray data sets spanning almost 30 years, and some have exposure times (for the quasi-quiescent time intervals) as short as $\approx$20\,min (for instance, the duration of the 2015 \emph{Swift} observations was only 28\,min). These complications make a more detailed analysis difficult.

\section{Conclusions}
\label{sect:summary}

We confirm that the magnetic activity of the K dwarf component in V471\,Tau is strongly influenced by the close white dwarf companion. Our related concluding remarks are as follows:
   \begin{itemize}
      \item We confirm that there is a permanent active longitude on the K star facing the white dwarf over decades, regardless of the actual phase of the magnetic cycle.
      \item We confirm a weak solar-type surface differential rotation on the surface of the K star, supporting theoretical expectations.
      \item We suggest that frequent flaring should have a significant contribution in heating the corona of the K star.
      \item From the long-term evolution of X-ray luminosity we confirm a magnetic activity cycle of $\approx$12.7\,ys, whose traces appear also in the chromospheric (H$\alpha$) activity pattern.
      \item We find that the inter-binary H$\alpha$ emission from the vicinity of the L1 point is correlated with the activity cycle: it intensifies during activity maximum and fades away during minimum.
   \end{itemize}

\begin{acknowledgements}
This publication makes use of VOSA, developed under the Spanish Virtual Observatory project supported by the Spanish MINECO through grant AyA2017-84089.
VOSA has been partially updated by using funding from the European Union's Horizon 2020 Research and Innovation Programme, under Grant Agreement No.\,776403 (EXOPLANETS-A).
This work was supported by the Hungarian National Research, Development and Innovation Office grant OTKA K131508, KH-130526, by the Lend\"ulet Program of the Hungarian Academy of Sciences, project No. LP2018-7/2019
and by the NKFIH grant 2019-2.1.11-TÉT-2019-00056. Authors acknowledge the financial support of the Austrian-Hungarian Action Foundation (98\"ou5, 101\"ou13).
TB acknowledges the financial support of the Hungarian National Research, Development and Innovation Office grant NKFIH KH-130372.
LK acknowledges the financial support of the Hungarian National Research, Development and Innovation Office grant NKFIH PD-134784.
\end{acknowledgements}

%\end{thebibliography}
\bibliography{v471tau}

\begin{appendix}

\twocolumn
\section{Log of spectroscopic data}\label{app1}

\vspace{0.6cm}
%---------------------------- TA1
\begin{center}
\tablehead{\hline\noalign{\smallskip}%
HJD & Date &  Phase & S/N & Subset\\
\hline\hline\noalign{\smallskip}}%
\topcaption{List of ESPaDOnS spectra of V471\,Tau taken from the CFHT Science Archive. Given are the Heliocentric Julian Dates (HJD$-$2\,450\,000), dates (dd.mm.yyyy), rotational phases computed using Eq.~\ref{eq:phase}, signal-to-noise ratios (S/N) for the four subsets (S1-S4) formed for Doppler imaging.}\label{table:cfht}
\begin{supertabular}{c c c r c}
\tabletail{\hline\noalign{\smallskip}}
\tablelasttail{\hline\noalign{\smallskip}}%
3718.714  & 14.12.2005   & 0.703 &  60  &  S1 \\ 
3718.717  & 14.12.2005   & 0.709 &  58  &  S1 \\ 
3718.720  & 14.12.2005   & 0.714 &  62  &  S1 \\ 
3718.723  & 14.12.2005   & 0.720 &  65  &  S1 \\ 
3718.728  & 14.12.2005   & 0.730 &  62  &  S1 \\ 
3718.731  & 14.12.2005   & 0.735 &  64  &  S1 \\ 
3718.734  & 14.12.2005   & 0.741 &  63  &  S1 \\ 
3718.736  & 14.12.2005   & 0.745 &  63  &  S1 \\ 
3718.740  & 14.12.2005   & 0.753 &  63  &  S1 \\ 
3718.742  & 14.12.2005   & 0.757 &  64  &  S1 \\ 
3718.745  & 14.12.2005   & 0.762 &  66  &  S1 \\ 
3718.748  & 14.12.2005   & 0.768 &  68  &  S1 \\ 
3718.751  & 14.12.2005   & 0.774 &  66  &  S1 \\ 
3718.754  & 14.12.2005   & 0.780 &  65  &  S1 \\ 
3718.757  & 14.12.2005   & 0.785 &  63  &  S1 \\ 
3718.760  & 14.12.2005   & 0.791 &  63  &  S1 \\ 
3718.763  & 14.12.2005   & 0.797 &  64  &  S1 \\ 
3718.766  & 14.12.2005   & 0.803 &  65  &  S1 \\ 
3718.769  & 14.12.2005   & 0.808 &  64  &  S1 \\ 
3718.772  & 14.12.2005   & 0.814 &  58  &  S1 \\ 
3718.776  & 14.12.2005   & 0.822 &  50  &  S1 \\ 
3718.779  & 14.12.2005   & 0.828 &  49  &  S1 \\ 
3718.782  & 14.12.2005   & 0.833 &  43  &  S1 \\ 
3718.785  & 14.12.2005   & 0.839 &  43  &  S1 \\ 
3718.788  & 14.12.2005   & 0.845 &  49  &  S1 \\ 
3718.791  & 14.12.2005   & 0.851 &  50  &  S1 \\ 
3718.794  & 14.12.2005   & 0.856 &  52  &  S1 \\ 
3718.797  & 14.12.2005   & 0.862 &  59  &  S1 \\ 
3718.800  & 14.12.2005   & 0.868 &  64  &  S1 \\ 
3718.803  & 14.12.2005   & 0.874 &  64  &  S1 \\ 
3718.806  & 14.12.2005   & 0.879 &  61  &  S1 \\ 
3718.808  & 14.12.2005   & 0.883 &  61  &  S1 \\ 
3718.811  & 14.12.2005   & 0.889 &  58  &  S1 \\ 
3718.814  & 14.12.2005   & 0.895 &  62  &  S1 \\ 
3718.817  & 14.12.2005   & 0.900 &  60  &  S1 \\ 
3718.820  & 14.12.2005   & 0.906 &  57  &  S1 \\ 
3718.825  & 14.12.2005   & 0.916 &  60  &  S1 \\ 
3718.828  & 14.12.2005   & 0.922 &  61  &  S1 \\ 
3718.831  & 14.12.2005   & 0.927 &  54  &  S1 \\ 
3718.833  & 14.12.2005   & 0.931 &  58  &  S1 \\ 
3718.837  & 14.12.2005   & 0.939 &  49  &  S1 \\ 
3718.849  & 14.12.2005   & 0.962 &  60  &  S1 \\ 
3718.852  & 14.12.2005   & 0.968 &  60  &  S1 \\ 
3718.855  & 14.12.2005   & 0.973 &  53  &  S1 \\ 
3718.858  & 14.12.2005   & 0.979 &  58  &  S1 \\ 
3718.861  & 14.12.2005   & 0.985 &  50  &  S1 \\ 
3718.864  & 14.12.2005   & 0.991 &  51  &  S1 \\ 
3718.867  & 14.12.2005   & 0.996 &  52  &  S1 \\ 
3718.870  & 14.12.2005   & 0.002 &  57  &  S1 \\ 
3718.873  & 14.12.2005   & 0.008 &  58  &  S1 \\ 
3718.876  & 14.12.2005   & 0.014 &  57  &  S1 \\ 
3718.879  & 14.12.2005   & 0.019 &  56  &  S1 \\ 
3718.881  & 14.12.2005   & 0.023 &  57  &  S1 \\ 
3718.884  & 14.12.2005   & 0.029 &  53  &  S1 \\ 
3718.887  & 14.12.2005   & 0.035 &  50  &  S1 \\ 
3718.890  & 14.12.2005   & 0.041 &  52  &  S1 \\ 
3718.893  & 14.12.2005   & 0.046 &  53  &  S1 \\ 
3718.904  & 14.12.2005   & 0.067 &  51  &  S1 \\ 
3718.907  & 14.12.2005   & 0.073 &  59  &  S1 \\ 
3718.909  & 14.12.2005   & 0.077 &  60  &  S1 \\ 
3718.912  & 14.12.2005   & 0.083 &  59  &  S1 \\ 
3718.915  & 14.12.2005   & 0.088 &  59  &  S1 \\ 
3718.918  & 14.12.2005   & 0.094 &  61  &  S1 \\ 
3718.921  & 14.12.2005   & 0.100 &  60  &  S1 \\ 
3718.924  & 14.12.2005   & 0.106 &  57  &  S1 \\ 
3718.927  & 14.12.2005   & 0.112 &  54  &  S1 \\ 
3718.930  & 14.12.2005   & 0.117 &  56  &  S1 \\ 
3718.933  & 14.12.2005   & 0.123 &  53  &  S1 \\ 
3718.936  & 14.12.2005   & 0.129 &  52  &  S1 \\ 
3718.939  & 14.12.2005   & 0.135 &  51  &  S1 \\ 
3718.942  & 14.12.2005   & 0.140 &  57  &  S1 \\ 
3718.945  & 14.12.2005   & 0.146 &  54  &  S1 \\ 
3718.948  & 14.12.2005   & 0.152 &  56  &  S1 \\ 
3718.951  & 14.12.2005   & 0.158 &  45  &  S1 \\ 
3718.954  & 14.12.2005   & 0.163 &  52  &  S1 \\ 
3718.957  & 14.12.2005   & 0.169 &  51  &  S1 \\ 
3718.960  & 14.12.2005   & 0.175 &  53  &  S1 \\ 
3718.963  & 14.12.2005   & 0.181 &  48  &  S1 \\ 
3718.966  & 14.12.2005   & 0.186 &  41  &  S1 \\ 
3718.969  & 14.12.2005   & 0.192 &  45  &  S1 \\ 
3718.972  & 14.12.2005   & 0.198 &  48  &  S1 \\ 
3718.975  & 14.12.2005   & 0.204 &  49  &  S1 \\ 
%3718.978  & 14.12.2005   & 0.209 &  31  &  S1 \\ 
%3718.981  & 14.12.2005   & 0.215 &  28  &  S1 \\ 
%3718.983  & 14.12.2005   & 0.219 &  27  &  S1 \\ 
%3718.986  & 14.12.2005   & 0.225 &  30  &  S1 \\ 
%3718.989  & 14.12.2005   & 0.230 &  25  &  S1 \\ 
%3718.992  & 14.12.2005   & 0.236 &  38  &  S1 \\ 
%3718.995  & 14.12.2005   & 0.242 &  43  &  S1 \\ 
3718.999  & 14.12.2005   & 0.250 &  43  &  S1 \\ 
3719.002  & 14.12.2005   & 0.255 &  40  &  S1 \\ 
3719.004  & 14.12.2005   & 0.259 &  43  &  S1 \\ 
3719.007  & 14.12.2005   & 0.265 &  52  &  S1 \\ 
3719.010  & 14.12.2005   & 0.271 &  46  &  S1 \\ 
3719.013  & 14.12.2005   & 0.277 &  45  &  S1 \\ 
3719.016  & 14.12.2005   & 0.282 &  46  &  S1 \\ 
3719.019  & 14.12.2005   & 0.288 &  42  &  S1 \\ 
%3719.023  & 14.12.2005   & 0.296 &  37  &  S1 \\ 
%3719.025  & 14.12.2005   & 0.300 &  38  &  S1 \\ 
%3719.028  & 14.12.2005   & 0.305 &  33  &  S1 \\ 
%3719.031  & 14.12.2005   & 0.311 &  23  &  S1 \\ 
%3719.035  & 14.12.2005   & 0.319 &  24  &  S1 \\ 
%3719.038  & 14.12.2005   & 0.324 &  27  &  S1 \\ 
%3719.041  & 14.12.2005   & 0.330 &  26  &  S1 \\ 
%3719.044  & 14.12.2005   & 0.336 &  24  &  S1 \\ 
3720.693  & 16.12.2005   & 0.500 &  56  &  S1 \\ 
3720.696  & 16.12.2005   & 0.506 &  59  &  S1 \\ 
3720.699  & 16.12.2005   & 0.511 &  63  &  S1 \\ 
3720.702  & 16.12.2005   & 0.517 &  64  &  S1 \\ 
3720.705  & 16.12.2005   & 0.523 &  64  &  S1 \\ 
3720.707  & 16.12.2005   & 0.527 &  70  &  S1 \\ 
3720.710  & 16.12.2005   & 0.533 &  68  &  S1 \\ 
3720.713  & 16.12.2005   & 0.538 &  69  &  S1 \\ 
3720.716  & 16.12.2005   & 0.544 &  69  &  S1 \\ 
3720.719  & 16.12.2005   & 0.550 &  70  &  S1 \\ 
3720.722  & 16.12.2005   & 0.556 &  68  &  S1 \\ 
3720.725  & 16.12.2005   & 0.561 &  69  &  S1 \\ 
3720.728  & 16.12.2005   & 0.567 &  73  &  S1 \\ 
3720.730  & 16.12.2005   & 0.571 &  73  &  S1 \\ 
3720.733  & 16.12.2005   & 0.577 &  71  &  S1 \\ 
3720.736  & 16.12.2005   & 0.582 &  71  &  S1 \\ 
3720.742  & 16.12.2005   & 0.594 &  71  &  S1 \\ 
3720.745  & 16.12.2005   & 0.600 &  72  &  S1 \\ 
3720.747  & 16.12.2005   & 0.604 &  72  &  S1 \\ 
3720.750  & 16.12.2005   & 0.609 &  74  &  S1 \\ 
3720.753  & 16.12.2005   & 0.615 &  74  &  S1 \\ 
3720.756  & 16.12.2005   & 0.621 &  75  &  S1 \\ 
3720.759  & 16.12.2005   & 0.627 &  74  &  S1 \\ 
3720.762  & 16.12.2005   & 0.632 &  74  &  S1 \\ 
3720.765  & 16.12.2005   & 0.638 &  76  &  S1 \\ 
3720.768  & 16.12.2005   & 0.644 &  76  &  S1 \\ 
3720.771  & 16.12.2005   & 0.650 &  76  &  S1 \\ 
3720.774  & 16.12.2005   & 0.655 &  76  &  S1 \\ 
3720.777  & 16.12.2005   & 0.661 &  76  &  S1 \\ 
3720.779  & 16.12.2005   & 0.665 &  76  &  S1 \\ 
3720.782  & 16.12.2005   & 0.671 &  77  &  S1 \\ 
3720.785  & 16.12.2005   & 0.676 &  77  &  S1 \\ 
3720.790  & 16.12.2005   & 0.686 &  78  &  S1 \\ 
3720.793  & 16.12.2005   & 0.692 &  78  &  S1 \\ 
3720.796  & 16.12.2005   & 0.698 &  77  &  S1 \\ 
3720.799  & 16.12.2005   & 0.703 &  77  &  S1 \\ 
3720.802  & 16.12.2005   & 0.709 &  78  &  S1 \\ 
3720.805  & 16.12.2005   & 0.715 &  78  &  S1 \\ 
3720.808  & 16.12.2005   & 0.721 &  80  &  S1 \\ 
3720.811  & 16.12.2005   & 0.726 &  80  &  S1 \\ 
3720.814  & 16.12.2005   & 0.732 &  79  &  S1 \\ 
3720.816  & 16.12.2005   & 0.736 &  79  &  S1 \\ 
3720.819  & 16.12.2005   & 0.742 &  78  &  S1 \\ 
3720.822  & 16.12.2005   & 0.747 &  78  &  S1 \\ 
3720.825  & 16.12.2005   & 0.753 &  55  &  S1 \\ 
3720.830  & 16.12.2005   & 0.763 &  77  &  S1 \\ 
3720.833  & 16.12.2005   & 0.769 &  79  &  S1 \\ 
3720.836  & 16.12.2005   & 0.774 &  79  &  S1 \\ 
3720.839  & 16.12.2005   & 0.780 &  80  &  S1 \\ 
3720.842  & 16.12.2005   & 0.786 &  79  &  S1 \\ 
3720.845  & 16.12.2005   & 0.792 &  77  &  S1 \\ 
3720.848  & 16.12.2005   & 0.797 &  75  &  S1 \\ 
3720.851  & 16.12.2005   & 0.803 &  74  &  S1 \\ 
3720.854  & 16.12.2005   & 0.809 &  74  &  S1 \\ 
3720.856  & 16.12.2005   & 0.813 &  72  &  S1 \\ 
3720.859  & 16.12.2005   & 0.818 &  70  &  S1 \\ 
3720.862  & 16.12.2005   & 0.824 &  70  &  S1 \\ 
3720.865  & 16.12.2005   & 0.830 &  71  &  S1 \\ 
3720.868  & 16.12.2005   & 0.836 &  71  &  S2 \\ 
3720.871  & 16.12.2005   & 0.841 &  71  &  S2 \\ 
3720.874  & 16.12.2005   & 0.847 &  72  &  S2 \\ 
3720.879  & 16.12.2005   & 0.857 &  70  &  S2 \\ 
3720.882  & 16.12.2005   & 0.863 &  69  &  S2 \\ 
3720.884  & 16.12.2005   & 0.866 &  69  &  S2 \\ 
3720.887  & 16.12.2005   & 0.872 &  68  &  S2 \\ 
3720.890  & 16.12.2005   & 0.878 &  66  &  S2 \\ 
3720.893  & 16.12.2005   & 0.884 &  66  &  S2 \\ 
3720.896  & 16.12.2005   & 0.889 &  65  &  S2 \\ 
3720.899  & 16.12.2005   & 0.895 &  66  &  S2 \\ 
3720.902  & 16.12.2005   & 0.901 &  64  &  S2 \\ 
3720.905  & 16.12.2005   & 0.907 &  63  &  S2 \\ 
3720.908  & 16.12.2005   & 0.912 &  64  &  S2 \\ 
3720.911  & 16.12.2005   & 0.918 &  63  &  S2 \\ 
3720.914  & 16.12.2005   & 0.924 &  64  &  S2 \\ 
3720.916  & 16.12.2005   & 0.928 &  64  &  S2 \\ 
3720.919  & 16.12.2005   & 0.934 &  64  &  S2 \\ 
3720.922  & 16.12.2005   & 0.939 &  67  &  S2 \\ 
3720.926  & 16.12.2005   & 0.947 &  66  &  S2 \\ 
3720.928  & 16.12.2005   & 0.951 &  63  &  S2 \\ 
3720.931  & 16.12.2005   & 0.957 &  62  &  S2 \\ 
3720.934  & 16.12.2005   & 0.962 &  65  &  S2 \\ 
3720.937  & 16.12.2005   & 0.968 &  68  &  S2 \\ 
3720.940  & 16.12.2005   & 0.974 &  69  &  S2 \\ 
3720.943  & 16.12.2005   & 0.980 &  68  &  S2 \\ 
3720.946  & 16.12.2005   & 0.985 &  66  &  S2 \\ 
3720.949  & 16.12.2005   & 0.991 &  65  &  S2 \\ 
3720.952  & 16.12.2005   & 0.997 &  62  &  S2 \\ 
3720.955  & 16.12.2005   & 0.003 &  63  &  S2 \\ 
3720.957  & 16.12.2005   & 0.006 &  62  &  S2 \\ 
3720.960  & 16.12.2005   & 0.012 &  60  &  S2 \\ 
3720.963  & 16.12.2005   & 0.018 &  61  &  S2 \\ 
3720.966  & 16.12.2005   & 0.024 &  61  &  S2 \\ 
3720.969  & 16.12.2005   & 0.030 &  63  &  S2 \\ 
3720.973  & 16.12.2005   & 0.037 &  64  &  S2 \\ 
3720.975  & 16.12.2005   & 0.041 &  60  &  S2 \\ 
3720.978  & 16.12.2005   & 0.047 &  62  &  S2 \\ 
3720.981  & 16.12.2005   & 0.053 &  60  &  S2 \\ 
3720.984  & 16.12.2005   & 0.058 &  63  &  S2 \\ 
3720.987  & 16.12.2005   & 0.064 &  60  &  S2 \\ 
3720.990  & 16.12.2005   & 0.070 &  60  &  S2 \\ 
3720.993  & 16.12.2005   & 0.076 &  62  &  S2 \\ 
3720.996  & 16.12.2005   & 0.081 &  61  &  S2 \\ 
3720.999  & 16.12.2005   & 0.087 &  59  &  S2 \\ 
3721.001  & 16.12.2005   & 0.091 &  57  &  S2 \\ 
3721.004  & 16.12.2005   & 0.097 &  58  &  S2 \\ 
3721.007  & 16.12.2005   & 0.102 &  57  &  S2 \\ 
3721.010  & 16.12.2005   & 0.108 &  53  &  S2 \\ 
3721.013  & 16.12.2005   & 0.114 &  54  &  S2 \\ 
3721.016  & 16.12.2005   & 0.120 &  52  &  S2 \\ 
3721.026  & 16.12.2005   & 0.139 &  53  &  S2 \\ 
%3721.029  & 16.12.2005   & 0.145 &  54  &  S2 \\ 
3721.032  & 16.12.2005   & 0.150 &  55  &  S2 \\ 
3721.035  & 16.12.2005   & 0.156 &  53  &  S2 \\ 
3721.038  & 16.12.2005   & 0.162 &  53  &  S2 \\ 
%3721.041  & 16.12.2005   & 0.168 &  52  &  S2 \\ 
%3721.043  & 16.12.2005   & 0.171 &  51  &  S2 \\ 
3721.046  & 16.12.2005   & 0.177 &  48  &  S2 \\ 
3722.683  & 18.12.2005   & 0.318 &  60  &  S2 \\ 
3722.686  & 18.12.2005   & 0.324 &  63  &  S2 \\ 
3722.689  & 18.12.2005   & 0.330 &  62  &  S2 \\ 
3722.692  & 18.12.2005   & 0.335 &  62  &  S2 \\ 
3722.695  & 18.12.2005   & 0.341 &  66  &  S2 \\ 
3722.698  & 18.12.2005   & 0.347 &  66  &  S2 \\ 
3722.701  & 18.12.2005   & 0.353 &  68  &  S2 \\ 
3722.703  & 18.12.2005   & 0.357 &  68  &  S2 \\ 
3722.708  & 18.12.2005   & 0.366 &  71  &  S2 \\ 
3722.711  & 18.12.2005   & 0.372 &  70  &  S2 \\ 
3722.714  & 18.12.2005   & 0.378 &  69  &  S2 \\ 
3722.717  & 18.12.2005   & 0.383 &  69  &  S2 \\ 
3722.720  & 18.12.2005   & 0.389 &  73  &  S2 \\ 
3722.723  & 18.12.2005   & 0.395 &  73  &  S2 \\ 
3722.725  & 18.12.2005   & 0.399 &  74  &  S2 \\ 
3722.728  & 18.12.2005   & 0.405 &  77  &  S2 \\ 
3722.731  & 18.12.2005   & 0.410 &  78  &  S2 \\ 
3722.734  & 18.12.2005   & 0.416 &  77  &  S2 \\ 
3722.737  & 18.12.2005   & 0.422 &  77  &  S2 \\ 
3722.740  & 18.12.2005   & 0.428 &  80  &  S2 \\ 
3722.743  & 18.12.2005   & 0.433 &  80  &  S2 \\ 
3722.746  & 18.12.2005   & 0.439 &  79  &  S2 \\ 
3722.749  & 18.12.2005   & 0.445 &  78  &  S2 \\ 
3722.752  & 18.12.2005   & 0.451 &  77  &  S2 \\ 
3722.755  & 18.12.2005   & 0.456 &  77  &  S2 \\ 
3722.758  & 18.12.2005   & 0.462 &  75  &  S2 \\ 
3722.761  & 18.12.2005   & 0.468 &  75  &  S2 \\ 
3722.764  & 18.12.2005   & 0.474 &  75  &  S2 \\ 
3722.767  & 18.12.2005   & 0.479 &  75  &  S2 \\ 
3722.770  & 18.12.2005   & 0.485 &  75  &  S2 \\ 
3722.772  & 18.12.2005   & 0.489 &  74  &  S2 \\ 
3722.775  & 18.12.2005   & 0.495 &  75  &  S2 \\ 
3722.778  & 18.12.2005   & 0.500 &  76  &  S2 \\ 
3722.781  & 18.12.2005   & 0.506 &  75  &  S2 \\ 
3722.784  & 18.12.2005   & 0.512 &  76  &  S2 \\ 
3722.787  & 18.12.2005   & 0.518 &  76  &  S2 \\ 
3722.790  & 18.12.2005   & 0.523 &  75  &  S2 \\ 
3722.793  & 18.12.2005   & 0.529 &  73  &  S2 \\ 
3722.796  & 18.12.2005   & 0.535 &  68  &  S2 \\ 
3722.799  & 18.12.2005   & 0.541 &  64  &  S2 \\ 
3722.805  & 18.12.2005   & 0.552 &  69  &  S2 \\ 
3722.808  & 18.12.2005   & 0.558 &  70  &  S2 \\ 
3722.811  & 18.12.2005   & 0.564 &  71  &  S2 \\ 
3722.814  & 18.12.2005   & 0.570 &  66  &  S2 \\ 
3722.817  & 18.12.2005   & 0.575 &  72  &  S2 \\ 
3722.819  & 18.12.2005   & 0.579 &  73  &  S2 \\ 
%3722.822  & 18.12.2005   & 0.585 &  32  &  S2 \\ 
3722.831  & 18.12.2005   & 0.602 &  77  &  S2 \\ 
3722.834  & 18.12.2005   & 0.608 &  77  &  S2 \\ 
3722.836  & 18.12.2005   & 0.612 &  78  &  S2 \\ 
3722.839  & 18.12.2005   & 0.617 &  77  &  S2 \\ 
3722.842  & 18.12.2005   & 0.623 &  77  &  S2 \\ 
3722.845  & 18.12.2005   & 0.629 &  75  &  S2 \\ 
3722.848  & 18.12.2005   & 0.635 &  76  &  S2 \\ 
3722.851  & 18.12.2005   & 0.641 &  78  &  S2 \\ 
3722.854  & 18.12.2005   & 0.646 &  77  &  S2 \\ 
3722.857  & 18.12.2005   & 0.652 &  78  &  S2 \\ 
3722.860  & 18.12.2005   & 0.658 &  78  &  S2 \\ 
3722.862  & 18.12.2005   & 0.662 &  77  &  S2 \\ 
3722.865  & 18.12.2005   & 0.667 &  79  &  S2 \\ 
3722.868  & 18.12.2005   & 0.673 &  78  &  S2 \\ 
3722.871  & 18.12.2005   & 0.679 &  80  &  S2 \\ 
3722.874  & 18.12.2005   & 0.685 &  79  &  S2 \\ 
3722.878  & 18.12.2005   & 0.692 &  80  &  S2 \\ 
3722.881  & 18.12.2005   & 0.698 &  79  &  S2 \\ 
3722.884  & 18.12.2005   & 0.704 &  80  &  S2 \\ 
3722.887  & 18.12.2005   & 0.710 &  80  &  S2 \\ 
3722.890  & 18.12.2005   & 0.715 &  80  &  S2 \\ 
3722.893  & 18.12.2005   & 0.721 &  80  &  S2 \\ 
3722.896  & 18.12.2005   & 0.727 &  81  &  S2 \\ 
3722.899  & 18.12.2005   & 0.733 &  80  &  S2 \\ 
3722.901  & 18.12.2005   & 0.736 &  80  &  S2 \\ 
3722.904  & 18.12.2005   & 0.742 &  80  &  S2 \\ 
3722.907  & 18.12.2005   & 0.748 &  76  &  S2 \\ 
3722.910  & 18.12.2005   & 0.754 &  78  &  S2 \\ 
3722.913  & 18.12.2005   & 0.759 &  79  &  S2 \\ 
3722.916  & 18.12.2005   & 0.765 &  77  &  S2 \\ 
3722.919  & 18.12.2005   & 0.771 &  76  &  S2 \\ 
3722.922  & 18.12.2005   & 0.777 &  76  &  S2 \\ 
3722.925  & 18.12.2005   & 0.783 &  76  &  S2 \\ 
3722.928  & 18.12.2005   & 0.788 &  76  &  S2 \\ 
3722.931  & 18.12.2005   & 0.794 &  76  &  S2 \\ 
3722.934  & 18.12.2005   & 0.800 &  76  &  S2 \\ 
3722.937  & 18.12.2005   & 0.806 &  77  &  S2 \\ 
3722.940  & 18.12.2005   & 0.811 &  77  &  S2 \\ 
3722.943  & 18.12.2005   & 0.817 &  75  &  S2 \\ 
3722.946  & 18.12.2005   & 0.823 &  75  &  S2 \\ 
3722.949  & 18.12.2005   & 0.829 &  76  &  S2 \\ 
3722.951  & 18.12.2005   & 0.832 &  75  &  S2 \\ 
3722.954  & 18.12.2005   & 0.838 &  75  &  S2 \\ 
3722.957  & 18.12.2005   & 0.844 &  75  &  S2 \\ 
3722.963  & 18.12.2005   & 0.855 &  75  &  S2 \\ 
3722.966  & 18.12.2005   & 0.861 &  74  &  S2 \\ 
3722.969  & 18.12.2005   & 0.867 &  73  &  S2 \\ 
3722.972  & 18.12.2005   & 0.873 &  75  &  S2 \\ 
3722.975  & 18.12.2005   & 0.878 &  75  &  S2 \\ 
3722.978  & 18.12.2005   & 0.884 &  74  &  S2 \\ 
3722.981  & 18.12.2005   & 0.890 &  74  &  S2 \\ 
3722.984  & 18.12.2005   & 0.896 &  74  &  S2 \\ 
3722.987  & 18.12.2005   & 0.901 &  75  &  S2 \\ 
3722.990  & 18.12.2005   & 0.907 &  75  &  S2 \\ 
3722.993  & 18.12.2005   & 0.913 &  75  &  S2 \\ 
3722.996  & 18.12.2005   & 0.919 &  75  &  S2 \\ 
3722.998  & 18.12.2005   & 0.923 &  74  &  S2 \\ 
3723.001  & 18.12.2005   & 0.928 &  74  &  S2 \\ 
3723.004  & 18.12.2005   & 0.934 &  75  &  S2 \\ 
3723.007  & 18.12.2005   & 0.940 &  75  &  S2 \\ 
3723.010  & 18.12.2005   & 0.946 &  74  &  S2 \\ 
3723.013  & 18.12.2005   & 0.951 &  74  &  S2 \\ 
3723.016  & 18.12.2005   & 0.957 &  74  &  S2 \\ 
3723.020  & 18.12.2005   & 0.965 &  73  &  S2 \\ 
3723.022  & 18.12.2005   & 0.969 &  72  &  S2 \\ 
3723.025  & 18.12.2005   & 0.974 &  65  &  S2 \\ 
3723.028  & 18.12.2005   & 0.980 &  70  &  S2 \\ 
%%%%%%%%%%%
7011.756 & 20.12.2014 & 0.097 & 100 & S3 \\
7011.760 & 20.12.2014 & 0.104 & 98 & S3 \\
7011.763 & 20.12.2014 & 0.110 & 101 & S3 \\
7011.766 & 20.12.2014 & 0.116 & 102 & S3 \\
7011.769 & 20.12.2014 & 0.122 & 102 & S3 \\
7011.772 & 20.12.2014 & 0.127 & 101 & S3 \\
7011.775 & 20.12.2014 & 0.133 & 100 & S3 \\
7011.778 & 20.12.2014 & 0.139 & 101 & S3 \\
7011.781 & 20.12.2014 & 0.145 & 101 & S3 \\
7011.784 & 20.12.2014 & 0.150 & 100 & S3 \\
7011.787 & 20.12.2014 & 0.156 & 98 & S3 \\
7011.790 & 20.12.2014 & 0.162 & 97 & S3 \\
%7011.998 & 20.12.2014 & 0.561 & 52 & S3 \\
%7012.001 & 20.12.2014 & 0.567 & 62 & S3 \\
%7012.004 & 20.12.2014 & 0.573 & 71 & S3 \\
7012.007 & 20.12.2014 & 0.578 & 71 & S3 \\
7012.010 & 20.12.2014 & 0.584 & 72 & S3 \\
7012.013 & 20.12.2014 & 0.590 & 66 & S3 \\
7012.016 & 20.12.2014 & 0.596 & 55 & S3 \\
7012.019 & 20.12.2014 & 0.601 & 55 & S3 \\
7012.022 & 20.12.2014 & 0.607 & 58 & S3 \\
7012.026 & 20.12.2014 & 0.615 & 57 & S3 \\
7012.029 & 20.12.2014 & 0.621 & 57 & S3 \\
7012.032 & 20.12.2014 & 0.626 & 47 & S3 \\
7012.732 & 21.12.2014 & 0.969 & 90 & S3 \\
7012.735 & 21.12.2014 & 0.975 & 91 & S3 \\
7012.738 & 21.12.2014 & 0.981 & 91 & S3 \\
7012.741 & 21.12.2014 & 0.987 & 93 & S3 \\
7012.744 & 21.12.2014 & 0.992 & 94 & S3 \\
7012.747 & 21.12.2014 & 0.998 & 94 & S3 \\
7012.750 & 21.12.2014 & 0.004 & 94 & S3 \\
7012.753 & 21.12.2014 & 0.010 & 94 & S3 \\
7012.756 & 21.12.2014 & 0.015 & 95 & S3 \\
7012.759 & 21.12.2014 & 0.021 & 93 & S3 \\
7012.762 & 21.12.2014 & 0.027 & 95 & S3 \\
7012.765 & 21.12.2014 & 0.033 & 95 & S3 \\
7012.973 & 21.12.2014 & 0.432 & 89 & S3 \\
7012.976 & 21.12.2014 & 0.438 & 87 & S3 \\
7012.979 & 21.12.2014 & 0.443 & 84 & S3 \\
7012.982 & 21.12.2014 & 0.449 & 88 & S3 \\
7012.986 & 21.12.2014 & 0.457 & 87 & S3 \\
7012.989 & 21.12.2014 & 0.462 & 88 & S3 \\
7012.992 & 21.12.2014 & 0.468 & 89 & S3 \\
7012.995 & 21.12.2014 & 0.474 & 83 & S3 \\
7012.998 & 21.12.2014 & 0.480 & 81 & S3 \\
7013.001 & 21.12.2014 & 0.486 & 83 & S3 \\
%7013.004 & 21.12.2014 & 0.491 & 77 & S3 \\
%7013.007 & 21.12.2014 & 0.497 & 75 & S3 \\
7013.772 & 22.12.2014 & 0.965 & 84 & S3 \\
7013.775 & 22.12.2014 & 0.971 & 89 & S3 \\
7013.778 & 22.12.2014 & 0.976 & 90 & S3 \\
7013.781 & 22.12.2014 & 0.982 & 90 & S3 \\
7013.784 & 22.12.2014 & 0.988 & 93 & S3 \\
7013.787 & 22.12.2014 & 0.994 & 95 & S3 \\
7013.791 & 22.12.2014 & 0.001 & 93 & S3 \\
7013.794 & 22.12.2014 & 0.007 & 95 & S3 \\
7013.797 & 22.12.2014 & 0.013 & 96 & S3 \\
7013.800 & 22.12.2014 & 0.019 & 96 & S3 \\
7013.803 & 22.12.2014 & 0.024 & 95 & S3 \\
7013.806 & 22.12.2014 & 0.030 & 92 & S3 \\
7014.013 & 22.12.2014 & 0.427 & 86 & S3 \\
7014.016 & 22.12.2014 & 0.433 & 83 & S3 \\
7014.019 & 22.12.2014 & 0.439 & 82 & S3 \\
7014.022 & 22.12.2014 & 0.445 & 81 & S3 \\
7014.025 & 22.12.2014 & 0.450 & 84 & S3 \\
7014.028 & 22.12.2014 & 0.456 & 84 & S3 \\
7014.031 & 22.12.2014 & 0.462 & 84 & S3 \\
7014.034 & 22.12.2014 & 0.468 & 81 & S3 \\
7014.038 & 22.12.2014 & 0.475 & 77 & S3 \\
7014.041 & 22.12.2014 & 0.481 & 75 & S3 \\
7014.044 & 22.12.2014 & 0.487 & 74 & S3 \\
%7014.047 & 22.12.2014 & 0.492 & 74 & S3 \\
7019.867 & 28.12.2014 & 0.659 & 98 & S3 \\
7019.870 & 28.12.2014 & 0.665 & 94 & S3 \\
7019.873 & 28.12.2014 & 0.671 & 99 & S3 \\
7019.876 & 28.12.2014 & 0.677 & 96 & S3 \\
7019.879 & 28.12.2014 & 0.682 & 98 & S3 \\
7019.882 & 28.12.2014 & 0.688 & 100 & S3 \\
7019.885 & 28.12.2014 & 0.694 & 99 & S3 \\
7019.889 & 28.12.2014 & 0.702 & 98 & S3 \\
7019.892 & 28.12.2014 & 0.707 & 85 & S3 \\
7019.895 & 28.12.2014 & 0.713 & 97 & S3 \\
7019.898 & 28.12.2014 & 0.719 & 97 & S3 \\
7019.901 & 28.12.2014 & 0.725 & 79 & S3 \\
7020.746 & 29.12.2014 & 0.346 & 97 & S3 \\
7020.749 & 29.12.2014 & 0.352 & 97 & S3 \\
7020.752 & 29.12.2014 & 0.357 & 96 & S3 \\
7020.755 & 29.12.2014 & 0.363 & 96 & S3 \\
7020.758 & 29.12.2014 & 0.369 & 96 & S3 \\
7020.761 & 29.12.2014 & 0.375 & 97 & S3 \\
7020.764 & 29.12.2014 & 0.380 & 95 & S3 \\
7020.767 & 29.12.2014 & 0.386 & 95 & S3 \\
7020.770 & 29.12.2014 & 0.392 & 94 & S3 \\
7020.773 & 29.12.2014 & 0.398 & 94 & S3 \\
7020.776 & 29.12.2014 & 0.403 & 93 & S3 \\
7020.780 & 29.12.2014 & 0.411 & 94 & S3 \\
7020.984 & 29.12.2014 & 0.803 & 92 & S3 \\
7020.988 & 29.12.2014 & 0.810 & 93 & S3 \\
7020.991 & 29.12.2014 & 0.816 & 93 & S3 \\
7020.994 & 29.12.2014 & 0.822 & 91 & S3 \\
7020.997 & 29.12.2014 & 0.828 & 91 & S3 \\
7021.000 & 29.12.2014 & 0.833 & 88 & S3 \\
7021.003 & 29.12.2014 & 0.839 & 88 & S3 \\
7021.006 & 29.12.2014 & 0.845 & 85 & S3 \\
7021.009 & 29.12.2014 & 0.851 & 86 & S3 \\
7021.012 & 29.12.2014 & 0.856 & 85 & S3 \\
7021.015 & 29.12.2014 & 0.862 & 86 & S3 \\
7021.018 & 29.12.2014 & 0.868 & 85 & S3 \\
7021.685 & 30.12.2014 & 0.148 & 95 & S3 \\
7021.688 & 30.12.2014 & 0.153 & 93 & S3 \\
7021.691 & 30.12.2014 & 0.159 & 94 & S3 \\
7021.694 & 30.12.2014 & 0.165 & 97 & S3 \\
7021.697 & 30.12.2014 & 0.171 & 97 & S3 \\
7021.703 & 30.12.2014 & 0.182 & 97 & S3 \\
7021.706 & 30.12.2014 & 0.188 & 96 & S3 \\
7021.709 & 30.12.2014 & 0.194 & 97 & S3 \\
7021.712 & 30.12.2014 & 0.199 & 99 & S3 \\
7021.715 & 30.12.2014 & 0.205 & 97 & S3 \\
7021.719 & 30.12.2014 & 0.213 & 99 & S3 \\
7021.928 & 30.12.2014 & 0.614 & 100 & S3 \\
7021.931 & 30.12.2014 & 0.620 & 101 & S3 \\
7021.934 & 30.12.2014 & 0.625 & 100 & S3 \\
7021.937 & 30.12.2014 & 0.631 & 100 & S3 \\
7021.940 & 30.12.2014 & 0.637 & 99 & S3 \\
7021.943 & 30.12.2014 & 0.643 & 101 & S3 \\
7021.946 & 30.12.2014 & 0.648 & 101 & S3 \\
7021.950 & 30.12.2014 & 0.656 & 100 & S3 \\
7021.953 & 30.12.2014 & 0.662 & 101 & S3 \\
7021.956 & 30.12.2014 & 0.668 & 99 & S3 \\
7021.959 & 30.12.2014 & 0.673 & 101 & S3 \\
7021.962 & 30.12.2014 & 0.679 & 101 & S3 \\
%%%%%%%%
7029.733 & 07.01.2015  & 0.589 & 60 & S4 \\
7029.736 & 07.01.2015  & 0.595 & 61 & S4 \\
7029.739 & 07.01.2015  & 0.601 & 62 & S4 \\
7029.742 & 07.01.2015  & 0.607 & 59 & S4 \\
7029.745 & 07.01.2015  & 0.612 & 61 & S4 \\
7029.748 & 07.01.2015  & 0.618 & 64 & S4 \\
7029.751 & 07.01.2015  & 0.624 & 62 & S4 \\
7029.755 & 07.01.2015  & 0.632 & 68 & S4 \\
7029.758 & 07.01.2015  & 0.637 & 71 & S4 \\
7029.761 & 07.01.2015  & 0.643 & 75 & S4 \\
7029.764 & 07.01.2015  & 0.649 & 82 & S4 \\
7029.767 & 07.01.2015  & 0.655 & 86 & S4 \\
7029.973 & 07.01.2015  & 0.050 & 80 & S4 \\
7029.976 & 07.01.2015  & 0.056 & 78 & S4 \\
7029.979 & 07.01.2015  & 0.061 & 75 & S4 \\
7029.982 & 07.01.2015  & 0.067 & 77 & S4 \\
7029.985 & 07.01.2015  & 0.073 & 74 & S4 \\
7029.988 & 07.01.2015  & 0.079 & 71 & S4 \\
7029.991 & 07.01.2015  & 0.084 & 71 & S4 \\
7029.994 & 07.01.2015  & 0.090 & 72 & S4 \\
7029.997 & 07.01.2015  & 0.096 & 75 & S4 \\
7030.000 & 07.01.2015  & 0.102 & 68 & S4 \\
7030.004 & 07.01.2015  & 0.109 & 71 & S4 \\
7030.007 & 07.01.2015  & 0.115 & 66 & S4 \\
7030.692 & 08.01.2015  & 0.429 & 98 & S4 \\
7030.695 & 08.01.2015  & 0.435 & 97 & S4 \\
7030.698 & 08.01.2015  & 0.441 & 98 & S4 \\
7030.701 & 08.01.2015  & 0.447 & 97 & S4 \\
7030.704 & 08.01.2015  & 0.452 & 97 & S4 \\
7030.707 & 08.01.2015  & 0.458 & 97 & S4 \\
7030.710 & 08.01.2015  & 0.464 & 97 & S4 \\
7030.713 & 08.01.2015  & 0.470 & 99 & S4 \\
7030.717 & 08.01.2015  & 0.477 & 97 & S4 \\
7030.720 & 08.01.2015  & 0.483 & 98 & S4 \\
7030.723 & 08.01.2015  & 0.489 & 98 & S4 \\
7030.726 & 08.01.2015  & 0.495 & 98 & S4 \\
7030.932 & 08.01.2015  & 0.890 & 88 & S4 \\
7030.935 & 08.01.2015  & 0.896 & 90 & S4 \\
7030.938 & 08.01.2015  & 0.901 & 89 & S4 \\
7030.941 & 08.01.2015  & 0.907 & 89 & S4 \\
7030.945 & 08.01.2015  & 0.915 & 90 & S4 \\
7030.948 & 08.01.2015  & 0.921 & 88 & S4 \\
7030.951 & 08.01.2015  & 0.926 & 88 & S4 \\
7030.954 & 08.01.2015  & 0.932 & 87 & S4 \\
7030.957 & 08.01.2015  & 0.938 & 87 & S4 \\
7030.960 & 08.01.2015  & 0.944 & 88 & S4 \\
7030.963 & 08.01.2015  & 0.949 & 88 & S4 \\
7030.966 & 08.01.2015  & 0.955 & 86 & S4 \\
7031.689 & 09.01.2015  & 0.342 & 95 & S4 \\
7031.692 & 09.01.2015  & 0.348 & 94 & S4 \\
7031.695 & 09.01.2015  & 0.354 & 94 & S4 \\
7031.698 & 09.01.2015  & 0.360 & 87 & S4 \\
7031.701 & 09.01.2015  & 0.365 & 92 & S4 \\
7031.704 & 09.01.2015  & 0.371 & 93 & S4 \\
7031.707 & 09.01.2015  & 0.377 & 91 & S4 \\
7031.710 & 09.01.2015  & 0.383 & 92 & S4 \\
7031.713 & 09.01.2015  & 0.388 & 91 & S4 \\
7031.716 & 09.01.2015  & 0.394 & 93 & S4 \\
7031.719 & 09.01.2015  & 0.400 & 94 & S4 \\
7031.722 & 09.01.2015  & 0.406 & 95 & S4 \\
7031.928 & 09.01.2015  & 0.801 & 90 & S4 \\
7031.931 & 09.01.2015  & 0.807 & 92 & S4 \\
7031.934 & 09.01.2015  & 0.812 & 86 & S4 \\
7031.937 & 09.01.2015  & 0.818 & 88 & S4 \\
7031.940 & 09.01.2015  & 0.824 & 89 & S4 \\
7031.943 & 09.01.2015  & 0.830 & 89 & S4 \\
7031.947 & 09.01.2015  & 0.837 & 85 & S4 \\
7031.950 & 09.01.2015  & 0.843 & 89 & S4 \\
7031.953 & 09.01.2015  & 0.849 & 83 & S4 \\
7031.956 & 09.01.2015  & 0.855 & 87 & S4 \\
7031.959 & 09.01.2015  & 0.860 & 81 & S4 \\
7031.962 & 09.01.2015  & 0.866 & 77 & S4 \\
7032.689 & 10.01.2015  & 0.261 & 86 & S4 \\
7032.692 & 10.01.2015  & 0.267 & 86 & S4 \\
7032.695 & 10.01.2015  & 0.273 & 88 & S4 \\
7032.698 & 10.01.2015  & 0.278 & 86 & S4 \\
7032.701 & 10.01.2015  & 0.284 & 85 & S4 \\
7032.705 & 10.01.2015  & 0.292 & 86 & S4 \\
7032.708 & 10.01.2015  & 0.298 & 85 & S4 \\
7032.711 & 10.01.2015  & 0.303 & 84 & S4 \\
7032.714 & 10.01.2015  & 0.309 & 84 & S4 \\
7032.717 & 10.01.2015  & 0.315 & 83 & S4 \\
7032.720 & 10.01.2015  & 0.321 & 89 & S4 \\
7032.723 & 10.01.2015  & 0.326 & 89 & S4 \\
7032.930 & 10.01.2015  & 0.723 & 89 & S4 \\
7032.933 & 10.01.2015  & 0.729 & 90 & S4 \\
7032.936 & 10.01.2015  & 0.735 & 86 & S4 \\
7032.939 & 10.01.2015  & 0.741 & 86 & S4 \\
7032.942 & 10.01.2015  & 0.747 & 87 & S4 \\
7032.945 & 10.01.2015  & 0.752 & 87 & S4 \\
7032.948 & 10.01.2015  & 0.758 & 84 & S4 \\
7032.951 & 10.01.2015  & 0.764 & 87 & S4 \\
7032.954 & 10.01.2015  & 0.770 & 87 & S4 \\
7032.957 & 10.01.2015  & 0.775 & 83 & S4 \\
7032.960 & 10.01.2015  & 0.781 & 82 & S4 \\
7032.963 & 10.01.2015  & 0.787 & 83 & S4 \\
7034.929 & 12.01.2015  & 0.559 & 91 & S4 \\
\end{supertabular}
\end{center}
%\end{multicols*}
%
\twocolumn

\begin{figure}[t]
        \centering
        \includegraphics[width=0.5\textwidth]{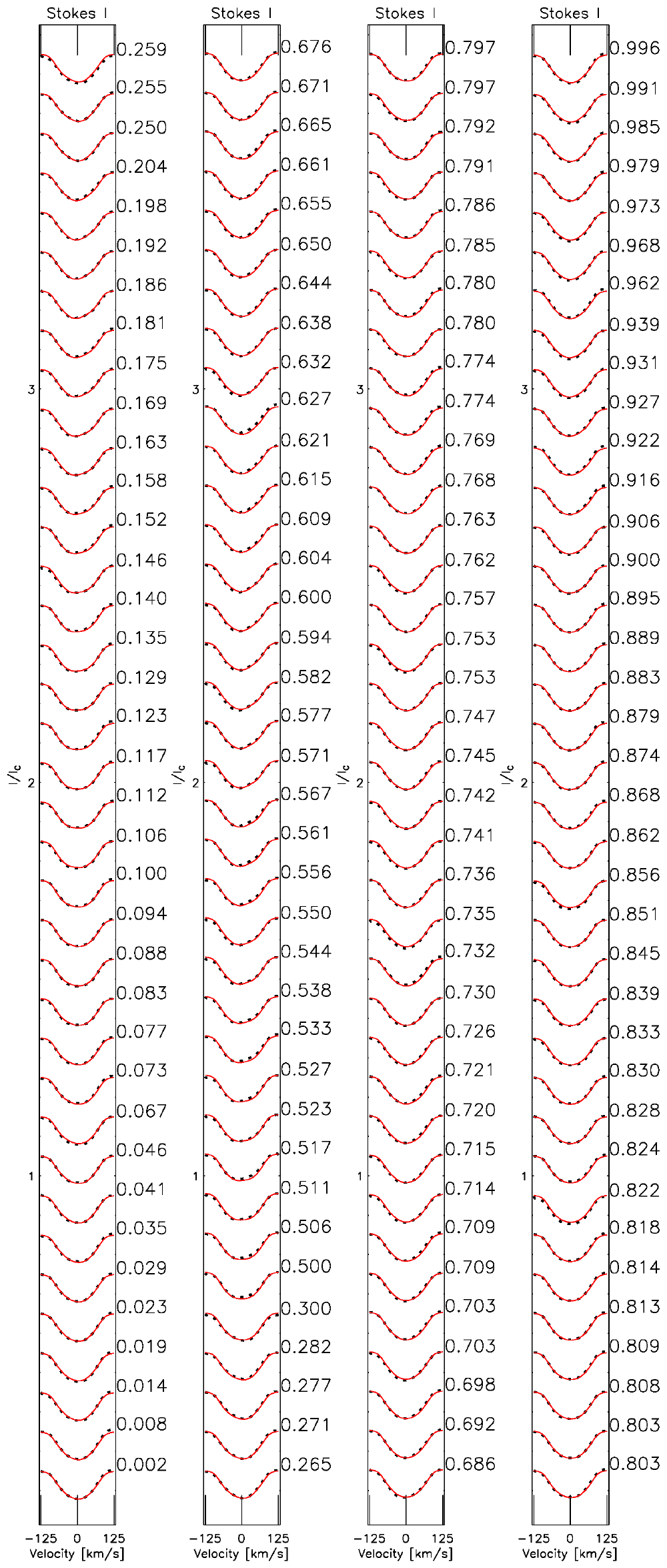}
      \caption{Observed line profiles (black dots) and their model fits (red lines) for the S1 Doppler reconstruction shown in Fig.\,\ref{fig_di12}. The phases of the individual observations are listed on the right side of the panels.
              }
         \label{fig:proffits_s1}
\end{figure}

\begin{figure}[t]
        \centering
        \includegraphics[width=0.5\textwidth]{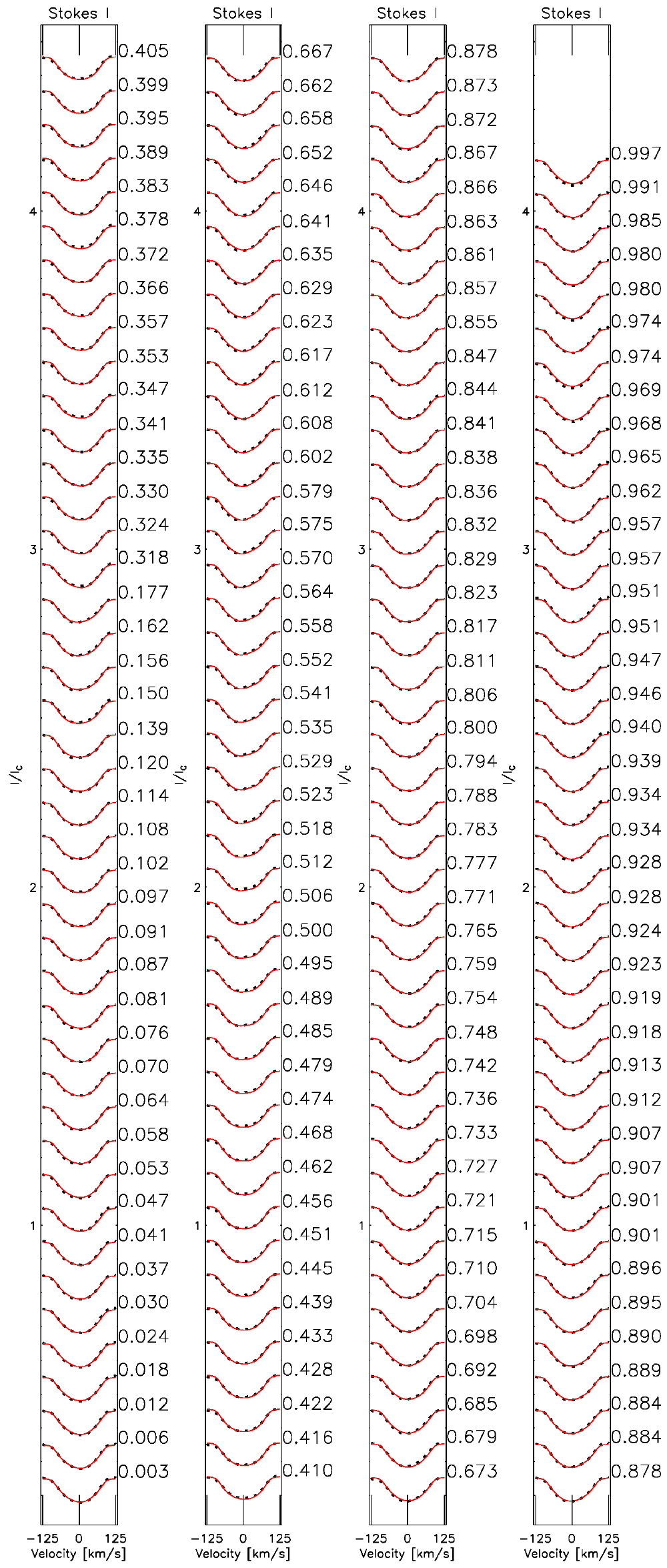}
      \caption{Fitted line profiles for the S2 Doppler reconstruction shown in Fig.\,\ref{fig_di12}. Otherwise as in Fig.\,\ref{fig:proffits_s1}.
              }
         \label{fig:proffits_s2}
\end{figure}

\begin{figure}[t]
        \centering
        \includegraphics[width=0.5\textwidth]{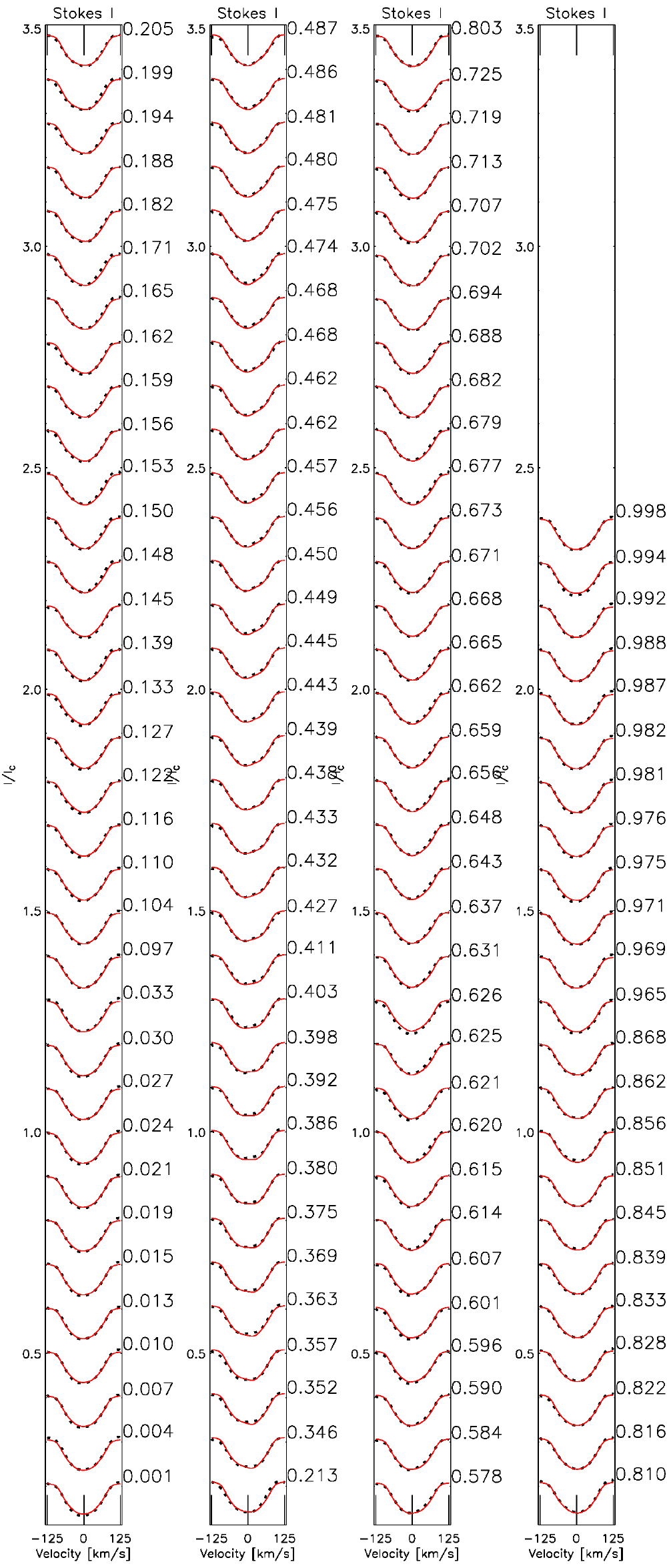}
      \caption{Fitted line profiles for the S3 Doppler reconstruction shown in Fig.\,\ref{fig_di34}. Otherwise as in Fig.\,\ref{fig:proffits_s1}.
              }
         \label{fig:proffits_s3}
\end{figure}

\begin{figure}[t]
        \centering
        \includegraphics[width=0.5\textwidth]{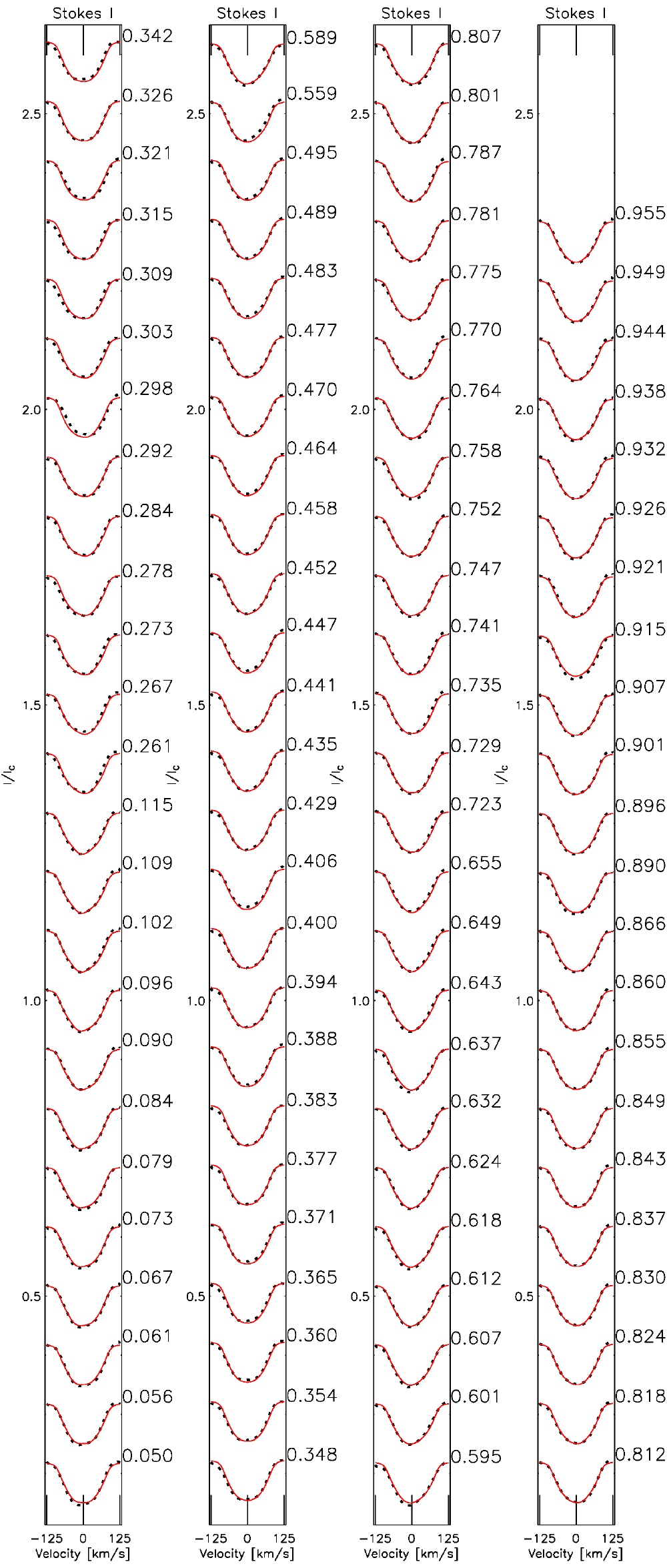}
      \caption{Fitted line profiles for the S4 Doppler reconstruction shown in Fig.\,\ref{fig_di34}. Otherwise as in Fig.\,\ref{fig:proffits_s1}.
              }
         \label{fig:proffits_s4}
\end{figure}

\end{appendix}

\end{document}